\documentclass{aa}

\usepackage{float} 

\usepackage{textcomp}
\usepackage[utf8]{inputenc}

\usepackage{amsmath, amssymb, amsfonts, amscd}
\usepackage{mathtools} 
\usepackage{array} 
\usepackage{cases} 
\usepackage{empheq} 
\usepackage{multirow} 
\usepackage{txfonts} 

\usepackage{graphicx} 
\usepackage{wrapfig} 

\usepackage[usenames,dvipsnames]{color} 
\usepackage{ulem} 
\usepackage[colorlinks=true,urlcolor=blue]{hyperref} 
\usepackage{natbib}
\newcommand{\Ms}{\ensuremath{M_\odot}}

\newcommand{\kms}{km~s$^{-1}$}

\usepackage[toc,page,titletoc]{appendix}
\usepackage{colortbl}

\begin{document}

\title{The coupling between internal waves and shear-induced turbulence in stellar radiation zones: the critical layers}

\author{L. Alvan \inst{1}, S. Mathis\inst{1}, T. Decressin\inst{2}}
\offprints{L. Alvan}

\institute{
Laboratoire AIM, CEA/DSM - CNRS - Universit\'e Paris Diderot, IRFU/SAp Centre de Saclay, F-91191 Gif-sur-Yvette, France
\and
Geneva Observatory, University of Geneva, chemin des Maillettes 51, CH-1290 Sauverny, Switzerland
 \\{}\\{}
 \email{lucie.alvan@cea.fr, stephane.mathis@cea.fr, thibaut.decressin@unige.ch} 
}

\date{}

\abstract
{Internal gravity waves (hereafter IGW) are known as {one of the} candidates for explaining the angular velocity profile in the Sun and in solar-type {main-sequence and evolved} stars, due to their role in the transport of angular momentum. Our bringing concerns critical layers, a process {poorly} explored in stellar physics, defined as the location where the local relative frequency of a given wave to the rotational frequency of the fluid tends to zero {({\it i. e.} that corresponds to co-rotation resonances)}.
}
{IGW propagate through stably-stratified radiative regions, where they extract or deposit angular momentum through two processes: radiative and viscous dampings and critical layers. Our goal is to obtain a complete picture of the effects of this latters.
}
{First, we expose a mathematical resolution of the equation of propagation for IGWs in adiabatic and non-adiabatic cases near critical layers. Then, the use of a dynamical stellar evolution code, which treats the secular transport of angular momentum, allows us to apply these results to the case of a solar-like star.}
{The analysis reveals two cases depending on the value of the Richardson number at critical layers: a stable one, where IGWs are attenuated as they pass through a critical level, and an unstable turbulent case where they can be reflected{/transmitted} by the critical level with a coefficient larger than one. Such over-reflection/transmission can have strong implications on our vision of angular momentum transport in stellar interiors.
}
{This paper highlights the existence of two regimes defining the interaction between an IGW and a critical layer. An application exposes the effect of the first regime, showing a strengthening of the damping of the wave. Moreover, this work opens new ways concerning the coupling between IGWs and shear instabilities in stellar interiors.
}


\titlerunning{Critical layers for internal waves in stellar radiation zones}
\authorrunning{L. Alvan, S. Mathis, T. Decressin}

\maketitle

\section{Introduction}

Thanks to helio- and asteroseismology, we are able to extract a huge amount of informations about solar and stellar structures and compositions ({\it e.g.} \citet{TCC2011,Aertsetal2010}) and their internal differential rotation profile \citep{Garciaetal2007,Becketal2012,Deheuvelsetal2012}. We know that internal rotation modifies the stellar structure since it generates flows, instabilities and chemical elements mixing, which modify stars evolution, for example their lifetime and their nucleosynthetic properties \citep[e.g.][and references therein]{Maeder2009}. Moreover, in order to understand the obtained solar/stellar rotation profiles and the related rotational history, it is essential to develop a complete theory incorporating the different angular momentum transport mechanisms occuring in stellar
interiors \citep[e.g.][and references therein]{Mathis2010}. In this context, one legitimely asks the question about the origin of the internal rotation profiles. 
Four main processes are responsible, in different ways, for the secular
angular momentum transport in radiative interiors.  First, a large-scale
meridional circulation is driven by structural
adjustments of stars, external applied torques and internal stresses
\citep[e.g.][]{Zahn1992, MathisZahn2004, DecressinMathisetal2009}. Next,
rotation profiles may be subject to different hydrodynamical shear and baroclinic instabilities and turbulence
\citep[e.g.][]{KS1982,TalonZahn1997,Maeder2003,Mathisetal2004}. Then, fossil magnetic fields, trapped during early-phases of stellar evolution
once radiation zones have been formed \citep{BS2004,DuezMathis2010}, can transport angular momentum through large-scale torques and Maxwell stresses
\citep[e.g.][]{GS1998,MathisZahn2005,GG2008,SBZ2011}. Finally, IGWs excited
at the convection/radiation boundaries constitute the fourth mechanism able
to transport angular momentum over large distances in stellar radiation
zones \citep[e.g.][]{Press1981,GN1989a,Schatzman1993,ZahnTalonMatias1997,TalonCharbonnel2005,MdB2012}. Note
that all these processes are not necessarily present at the same time
everywhere in the H.-R. diagram. Moreover, they act on various
characteristic timescales in stars of different masses and ages.\\ 

The object of this paper is the propagation of IGWs and the way they
interact with the shear (i.e. the differential rotation) of the surrounding
fluid. They are common in the terrestrial atmosphere and oceans
\citep[][]{Eckart1961,ChapmanLindzen1970}, that is why they are pretty well
known in Geophysics. We will use this advantage for their study in the
stellar case. We here draw attention to the mechanism whereby IGWs exchange
energy with the mean flow, independently from other dissipative processes
such as thermal and viscous diffusion.
{Indeed}, when the frequency of excited waves is of the same order as the angular velocity of the fluid (we will see the accurate definition later), a phenomenon of resonance occurs, which affects the properties of both the wave and the shear of the surrounding fluid. This phenomenon is called a critical layer. Under the assumption of a perfect fluid (neither heat conductor nor viscous), \cite{BookerBretherton1967} and \cite{LindzenBarker1985} have provided first results about critical layers in the geophysical case. They have shown that, depending on the value of the Richardson number of the fluid, which compares the relative strength of the shear and of the stable stratification (see Eq. \ref{eq:Ri}), the waves might be either attenuated or reflected by the critical layer. This reflection may even be an over-reflection together with an over-transmission (see also \cite{SutherlanYewchuk2004} for a laboratory evidence). In the same time, \cite{Koppel1964, Hazel1967, BaldwinRoberts1970, VanDuinKelder1986} have completed this
work in taking into account the conduction of heat and the viscosity of the
fluid. Surprinsingly, their conclusions about the role of critical layers
are identical. However, all these authors have produced their study in
cartesian coordinates, assuming that the stiffness of the domain was thin
enough to neglect its curvature. In the case of stellar radiation
zones where critical layers may play an important role
\citep[e.g.][]{BarkerOgilvie2010,Barker2011,Rogersetal2012}, these
equations should thus be generalized to the case of spherical coordinates
to be able to treat deep spherical shells.\\

Therefore, after exposing our notations and assumptions (\S 2.), we present a complete mathematical study of critical layers in the case of a perfect fluid (\S 3.) and of a non-perfect fluid (\S 4.)  in spherical geometry. Then, we determine the related mean vertical flux of angular momentum transported by IGWs (\S 5.). Next, we implement our theoretical results in the dynamical stellar evolution code STAREVOL and we apply our formalism to the evolution of a one solar-mass star (\S 6.). Finally, we present the conclusion and perspectives of this work (\S 7.).

\section{Definition, notations and hypotheses}

The star we consider is composed (at least) with a convective and a radiative region. IGWs are excited at the boundary between these two regions. They propagate in the radiative zone and are evanescent in the convective zone \citep[e.g.][]{Press1981}. As explained in the introduction, in Geophysics, the study of IGWs is usually made in 
cartesian coordinates, justified by a local approach. In the stellar case however, a global approach using the spherical coordinates ($r$,$\theta$,$\varphi$) is necessary.\\
 
In the frame of \cite{Zahn1992}, we choose a shelular angular velocity for the studied star's radiation zone: $\Omega(r,\theta) = \overline{\Omega}(r)$, considering that,
because of the strong stable stratification, the shear instability
decreases the horizontal gradient of the angular velocity
\citep[e.g.][]{TalonZahn1997,Maeder2003,Mathisetal2004}, which consequently
can be considered as only dependent of radius. For the moment, we neglect
the action of Coriolis and centrifugal accelerations while the Doppler
shift due to differential rotation is retained. We also neglect the action
of a potential magnetic field.\\ 

Now, we need to introduce some quantities to describe the properties of the fluid in the studied radiative zone. Each scalar field $X$ is written as
\begin{equation}
X\left(r,\theta,\varphi,t\right)={\overline X}\left(r\right)+X'\left(r,\theta,\varphi,t\right),
\label{expan} 
\end{equation}
where we have introduced its horizontal average on an isobar, $\overline X$, and its associated fluctuation $X'$. The thermodynamic variables employed are the density $\rho$, the pressure $p$, the temperature $T$ and the specific entropy $S$. {Next,} the stratification is described in terms of the Brunt-Va\"is\"al\"a frequency
\begin{equation}
\label{eq:BVfreq}
N^2 =
-\bar g\left(\frac{1}{\bar\rho} \frac{\partial \bar\rho}{\partial 
  r}-\frac{1}{\Gamma_1\bar{p}}\frac{\partial \bar{p}}{\partial r}\right)\hbox{,}
\end{equation}
where $\bar g(r)$ is the mean gravity in the Cowling approximation
where fluctuations of the gravific potential are neglected \citep[see][]{Cowling1941}, and
$\Gamma_1=\left(\displaystyle\frac{\partial \ln \bar p}{\partial \ln{\bar\rho}}\right)$
the adiabatic exponent. The relative importance of the stable stratification restoring force and the shear destabilizing effects is quantified
thanks to the Richardson number 
\begin{equation}
\label{eq:Ri}
\mathrm{Ri}=\frac{N^2}{\left(r\displaystyle\frac{\mathrm d\overline\Omega}{\mathrm d r}\right)^2}\hbox{.}
\end{equation}
When $\mathrm{Ri}$ is small, the velocity shear overcomes the stabilizing buoyancy and turbulence and mixing occur \citep[e.g.][]{TalonZahn1997}. On the contrary, when $\mathrm{Ri}$ is large, the fluid remains stable. Finally, for the case of a non-perfect fluid (\S 4.), we introduce the viscosity $\nu$ of the fluid and the coefficient of thermal conductivity $\kappa$. These notations will be recall at the proper moment.\\

IGWs themselves are characterized with their relative frequency
\begin{equation}
\sigma(r)=\sigma_{w}+m\Delta\Omega(r)\hbox{,}
\label{Doppler}
\end{equation}
where $\sigma_{w}$ is their excitation frequency (from the base of the convective zone in low-mass stars or the top of the convective core in intermediate and high mass stars). $m$ corresponds to a Fourier expansion along the longitudinal direction (c.f. Eq \ref{eq:5}) and $\Delta\Omega(r)=\overline\Omega(r)-\Omega_{CZ}$ is the difference between the
angular velocity at the level $r$ and at the  border with the convective zone. The introduction of $m$ leads to define two classes of waves. Prograde (respectively retrograde) waves correspond to negative (respectively positive) values of $m$.\\

Finally, all these notations allow us to define properly a critical
layer. It arises for a wave whose relative frequency $\sigma(r)$ becomes
zero. The coherence with the qualitative definition given in the
introduction is respected since $\sigma(r_c)=0$ means that the excitation
frequency of the wave equals $-m$ times the angular velocity of the
fluid. It corresponds to a corotation resonance. It is important to
highlight that the position of the critical layer depends on both wave
caracteristics ($m$ and $\sigma_{w}$) and shear properties
($\Delta\Omega(r)$). In our approximation, $\sigma(r_c)=0$ means that
$\overline\Omega(r_c)$ is constant so the critical levels are isobaric surfaces of the star.

\section{Case of the perfect fluid}
In this first approach, we treat the hydrodynamic equations assuming that the fluid is neither viscous nor heat conductor. 

\subsection{Equation of propagation of IGW near a critical layer}

Our aim is to calculate the Eulerian velocity field of a fluid particle. We introduce the time $t$ and the unit vectors ($\vec{\hat e}_r,\vec{\hat e}_\theta,\vec{\hat e}_\varphi$), associated with the classical spherical coordinates ($r$,$\theta$,$\varphi$). The velocity field is
\begin{equation}
\vec{V}(r,\theta, \varphi,t) = r
\sin\theta \bar\Omega(r)\vec{\hat e}_\varphi + \vec{u}(r,\theta,\varphi,t) \hbox{,}
\end{equation}
where $\vec{u}$ is the velocity associated with the wave and
$V_{\varphi}^{\Omega}=r\sin\theta\bar\Omega(r)$ the azimuthal velocity
field of the differential rotation. Note that we have neglected all other large scale velocities such as meridionnal circulation. Then, we give the linearized equations of hydrodynamics governing IGWs dynamics in an inertial frame, i.e. the momentum, continuity and energy equations:
\begin{equation}
\label{Eq:HydroP}
  \left\{
      \begin{array}{ll}
D_t\vec{u}=-\displaystyle\frac{\vec{\nabla}p'}{\bar\rho}+\displaystyle\frac{\rho'}{\bar\rho}\vec{g}
 \hbox{,}\\ 
D_t\rho'+\vec{\nabla}.(\bar\rho\vec{u})=0
 \hbox{,}\\ 
D_t\left(\displaystyle\frac{\rho'}{\bar\rho}-\frac{1}{\Gamma_1}\displaystyle\frac{p'}{\bar p}\right)-
\displaystyle\frac{N^2}{\bar{g}}u_r=0
\,\hbox{,}
      \end{array}
    \right.
\end{equation}
where $D_t = \partial_t +\Delta\Omega \partial_{\varphi}$. The equation of energy conservation is obtained using the linearized equation of state 
\begin{equation}
  \label{eq:State}
  \frac{\rho'}{\bar\rho}=\frac{p'}{\bar p}-\frac{T'}{\bar
    T}=\frac{p'}{\Gamma_1 \bar p}-\frac{S'}{c_P} \hbox{,}
\end{equation}
and assuming the ideal gaz law
\begin{equation}
  \label{eq:GazLaw}
  \bar p = \mathcal{R} \bar\rho \bar T \hbox{,}
\end{equation}
where $\mathcal{R}$ is the gaz constant and $c_P$ the specific heat per unit mass at constant pressure.\\

We define the Lagrangien displacement $\vec{\xi}$ as $\vec{u} = D_t\vec{\xi}$ and following \cite{Rieutord:1986}, we expand it on the vectorial spherical harmonics basis ($\vec R_l^m,\vec S_l^m,\vec T_l^m$) defined by:
\begin{equation}
\label{Eqy_lm}
  \left\{
      \begin{array}{l}
       \vec R_l^m\left(\theta,\varphi\right) = Y_l^m \left(\theta,\varphi\right)\vec{\hat e}_r \hbox{,}\\
         \vec{S}_l^m\left(\theta,\varphi\right) = \vec\nabla_\bot Y_l^m  = \partial_\theta Y_l^m \vec{\hat e}_\theta
+ \displaystyle\frac{1}{\sin\theta}\partial_\varphi Y_l^m \vec{\hat e}_\varphi \hbox{,}\\
  \vec{T}_l^m\left(\theta,\varphi\right) =  \vec\nabla_\bot \times \vec R_l^m =
  \displaystyle\frac{1}{\sin\theta}\partial_\varphi Y_l^m \vec{\hat e}_\theta - \partial_\theta
  Y_l^m \vec{\hat e}_\varphi \hbox{,}
      \end{array}
    \right.
\end{equation}
where $Y_l^m$ are the spherical harmonics with $l\in\mathbb{N}$ and $m \in \llbracket -l,l \rrbracket$. Thus: 
\begin{eqnarray}
  \label{eq:5}
  \vec\xi (r,\theta,\varphi,t)=
 && \sum_{l=0}^{\infty}\sum_{m=-l}^{l}\left\{
  \hat{\xi}_{r;l,m} (r)\vec R_l^m(\theta,\varphi)+\right.\\ \nonumber
&&{\left. \hat{\xi}_{H;l,m} (r)\vec S_l^m(\theta,\varphi)+
  \hat{\xi}_{T;l,m} (r)\vec T_l^m(\theta,\varphi)
\right\}}e^{i\sigma_{w}t}\hbox{.}
\end{eqnarray}
Then, we decompose $\rho'$ and $p'$ using spherical harmonics: 
\begin{equation}
\rho'(r,\theta,\varphi,t)=\displaystyle\sum_{l=0}^{\infty}\sum_{m=-l}^{l}\left\{\hat{\rho}'_{l,m}(r) Y_{l,m}(\theta,\varphi)\right\}e^{i\sigma_{w}
          t}\hbox{,}
\end{equation}
\begin{equation}
p'(r,\theta,\varphi,t)=\displaystyle\sum_{l=0}^{\infty}\sum_{m=-l}^{l}\left\{\hat{p}'_{l,m}(r) Y_{l,m}(\theta,\varphi)\right\}e^{i\sigma_{w}
          t}\hbox{,}
\end{equation}
and we obtain a new system made up of {radial} equations of momentum
\begin{equation}
\left\{
\begin{array}{l}
\bar\rho \sigma^2 \hat\xi_{r;l,m}=\displaystyle\frac{\mathrm d
  \hat{p}'_{l,m}}{dr}+\hat\rho'_{l,m}\bar g \hbox{,}\\ \nonumber
\bar\rho \sigma^2 \hat\xi_{H;l,m}=\displaystyle\frac{\hat{p}'_{l,m}}{r} \hbox{,}\\
\bar\rho \sigma^2 \hat\xi_{T;l,m}=0 \hbox{,}
\end{array}
\right.
\end{equation}
of mass conservation
\begin{equation}
\hat\rho'_{l,m}+\frac{1}{r^2} \frac{\partial}{\partial r}(r^2 \bar\rho \hat\xi_{r;l,m})
-\displaystyle\frac{l(l+1)}{r}\bar\rho\hat\xi_{H,l,m} = 0
 \hbox{,}
\end{equation}
and of energy in the adiabatic limit
\begin{equation}
\frac{\hat\rho'_{l,m}}{\bar\rho}=\frac{1}{\Gamma_1}\frac{\hat{p}'_{l,m}}{\bar{p}}+\frac{N^2}{\bar
  g}\hat\xi_{r;l,m}
\hbox{.}
\end{equation}

The combination of these relations leads to the system presented by \cite{Press1981}:
\begin{equation}
\label{Sysp'}
  \left\{
      \begin{array}{l}
        \displaystyle\frac{\mathrm d y_{l,m}}{\mathrm dr}=\left(\sigma^2-N^2\right)\hat\xi_{r;l,m}+ 
        \frac{N^2}{\bar g}y_{l,m} \hbox{,}\\
         \displaystyle\frac{\mathrm d}{\mathrm dr}\left(r^2\hat\xi_{r;l,m}\right)+\displaystyle\frac{1}{\Gamma_1}\frac{\mathrm
         d \ln \bar p}{\mathrm dr}\left(r^2\hat\xi_{r;l,m}\right)=\left[\displaystyle\frac{l(l+1)}{\sigma^2}-\frac{\bar
        \rho}{\Gamma_1 \bar p}r^2\right]
       y_{l,m}\hbox{,}
      \end{array}
    \right.
\end{equation}
where $y_{l,m}(r)=\hat{p}'_{l,m}/\bar{\rho}$.

In stellar radiative regions, the transport of angular momentum is dominated by low-frequency IGWs with $\sigma \ll N$, where $N$ is the Brunt-Va\"is\"al\"a frequency defined in Eq. (\ref{eq:BVfreq}). It allows us to apply the anelastic approximation \citep{Press1981} {where acoustic waves are filtered out}. Introducing
\begin{equation}
  \label{eq:9}
  \Psi_{l,m}(r)=\bar\rho^{\frac{1}{2}}r^2\hat\xi_{r;l,m}\hbox{,}
\end{equation}
we obtain the following equation of propagation: 
\begin{eqnarray}
  \label{eq:7}
\frac{\mathrm d^2 \Psi_{l,m}}{\mathrm dr^2}
&+&\left(\frac{N^2}{\sigma^2}-1\right)\frac{l(l+1)}{r^2} \Psi_{l,m}\\ \nonumber
&=&\left[\frac{1}{4}\left(\frac{\mathrm d \ln\bar\rho}{\mathrm d
      r}\right)^2+\frac{1}{2}\frac{\mathrm d^2 \ln\bar\rho}{\mathrm d
    r^2}-\frac{1}{\Gamma_1}\frac{\mathrm d^2 \ln \bar p}{\mathrm d r^2}\right] \Psi_{l,m}
 \hbox{.} 
\end{eqnarray}
As the right-hand side of Eq. \eqref{eq:7} is of order $1/H_\mathrm{P}^2$ with $H_\mathrm{P}$ the characteristic pressure or density height scale, it can be neglected if
\begin{equation}
  \label{eq:8}
  \left(\frac{N^2}{\sigma^2}-1\right)\frac{l(l+1)}{r^2} \gg \frac{1}{H_\mathrm{P}^2} \hbox{,}
\end{equation}
which is the case here. Finally, we obtain the equation of propagation of IGWs in a perfect fluid :
\begin{equation}
\label{eq:TGSspherical}
\frac{\mathrm d^2 \Psi_{l,m}}{\mathrm dr^2} + k_V^2(r) \Psi_{l,m}=0 \hbox{,}
\end{equation}
with
\begin{equation}
  \label{eq:10}
  k_V^2(r) = \left(\frac{N^2}{\sigma^2}-1\right)\frac{l(l+1)}{r^2} \hbox{.}
\end{equation}
In cartesian coordinates, this equation is called the Taylor, Goldstein and
Synge equation (TGS). We observe that the value $r=r_c$, where
$\sigma(r_c)=0$, is a singular point for this equation. Thus, we now
focus onto the study of the behavior of this equation around such a critical point.
Then for $r\approx r_c$, the equation of propagation becomes 
\begin{equation}
\label{Eq:TGS}
\frac{\mathrm d^2 \Psi_{l,m}}{\mathrm dr^2} +
\left[\frac{l(l+1)}{m^2}\frac{\mathrm{Ri}_c}{(r-r_c)^2}-k_{Hc}^2\right]\Psi_{l,m}=0 \hbox{,}
\end{equation}
where 
\begin{equation}
\mathrm{Ri}_c=\left(\frac{N^2}{\left(r\displaystyle\frac{\mathrm d\bar\Omega}{\mathrm d r}\right)^2}\right)_{r=r_c}\hbox{,}
\end{equation}
 is the value of the Richardson number at the critical level, and
\begin{equation}
\label{Eq:khc}
k_{Hc}=\displaystyle\frac{\sqrt{l(l+1)}}{r_c}\hbox{,}
\end{equation}
the horizontal wavenumber at the critical layer. We saw in the introduction (Eq. (\ref{eq:Ri})) that the Richardson number is relevant to distinguish between the relative importance of shear and stratification. We describe this divergence of cases from a quantitave point of view in the following part. 

\subsection{Mathematical resolution with the method of Frobenius}
The Frobenius method offers an infinite series solution for
a second-order ordinary differential equation of the form
\begin{eqnarray*}
  \label{eq:11}
  u''+p(z)u'+q(z)u=0 \hbox{,}\\
 \hbox{where }
  p(z)=\frac{1}{z}\sum_{j=0}^{\infty} p_j z^j \hbox{and }
  q(z)=\frac{1}{z^2}\sum_{j=0}^{\infty} q_j z^j
\end{eqnarray*}
in the vicinity of the singular point $z_0$=0. For more details, a mathematical description of this method can be found in \cite{Teschl}.\\
In our case, $p(r)=0$ and $q(r)=\frac{l(l+1)}{m^2}\frac{\mathrm{Ri}_c}{(r-r_c)^2}-k_{Hc}^2$ and the theory brings out two cases depending on the value of $l$, $m$ and $\mathrm{Ri}_c$: 

\begin{itemize}
\item Case 1: $\displaystyle\frac{l(l+1)}{m^2}\mathrm{Ri}_c =
  \displaystyle\frac{1}{4}$ \\
The solution is
\begin{equation}
\Psi_{l,m}^{\rm Fro}(r) = A_1(r-r_c)^{1/2} + B_1 (r-r_c)^{1/2}\log(|r-r_c|)\hbox{,}
\end{equation}
where $A_1$ and $B_1$ are constants representing the wave amplitude.\\

\item Case 2: $\displaystyle\frac{l(l+1)}{m^2}\mathrm{Ri}_c \ne \displaystyle\frac{1}{4}$\\
We define the complex parameter
\begin{equation}
\label{eq:etalm}
\eta_{l,m}=\sqrt{\frac{1}{4}-\frac{l(l+1)}{m^2}\mathrm{Ri}_c} \hbox{,}
\end{equation} 
and the solution is given by
\begin{equation}
\label{Eq:case2}
\Psi_{l,m}^{\rm Fro}(r) = A_2 (r-r_c)^{1/2+\eta_{l,m}} + B_2 (r-r_c)^{1/2-\eta_{l,m}}\hbox{,}
\end{equation}
where $A_2$ and $B_2$ are constants.

\end{itemize}

We exclude this special value of $\mathrm{Ri}_c$ and choose to consider only two different cases for the rest of the paper: $\frac{l(l+1)}{m^2}\mathrm{Ri}_c >\frac{1}{4}$ and $\frac{l(l+1)}{m^2}\mathrm{Ri}_c <
\frac{1}{4}$. The results can then be extended to the case $\frac{l(l+1)}{m^2}\mathrm{Ri}_c = \frac{1}{4}$ without recourse to the logarithmic solution \citep{VanDuinKelder1986}.\\

Let us now discuss the hydrodynamical behavior corresponding to the situations where $\mathrm{Ri}_c >\frac{1}{4}\frac{m^2}{l\left(l+1\right)}$ and $\mathrm{Ri}_c <
\frac{1}{4}\frac{m^2}{l\left(l+1\right)}$. Applying the classical method
exposed in \cite{DrazinReid2004} to the generalized spherical
Taylor-Goldstein-Synge equation (Eq. \ref{Eq:TGS}), we can identify that the first regime
corresponds to the case where the fluid stays stable with respect to the
vertical shear instability at the critical layer. {In the other case},
these instability and thus turbulence can develop. This clearly shows how
this is necessary to go beyond the current used formalisms for stellar
evolution where IGWs and vertical shear instability are considered as
uncoupled. Indeed, if a fluid becomes shear unstable, mixing occurs that
modify the local stratification and thus IGWs propagation \citep[see
e.g.][]{BrownSutherland2006,NaultSutherland2007}. Thus, we will now
distinguish the case of critical layers when the fluid is stable from the
one when it is unstable.

\subsection{The case of stable critical layers: $\mathrm{Ri}_c > \frac{1}{4}\frac{m^2}{l(l+1)}$}
It is now time to understand the physical behaviour of the solutions given above. First of all, we propose to cut the physical domain into two parts: above and below the
critical layer. As $\frac{l(l+1)}{m^2}\mathrm{Ri}_c >\frac{1}{4}$, $\eta_{l,m}$ defined by \eqref{eq:etalm} is a purely imaginary number. In order to clearly distinguish between real and imaginary parts, we introduce
\begin{equation}
\label{eq:alphalm}
\alpha_{l,m}=\sqrt{\frac{l(l+1)}{m^2}\mathrm{Ri}_c-\frac{1}{4}}=i\eta_{l,m} \hbox{.}
\end{equation} 
The two solutions can be written as
\begin{equation}
  \left\{
      \begin{array}{l}
\Psi^{\rm Fro}_{l,m+}(r) = A_{+} (r-r_c)^{1/2+i\alpha_{l,m}} 
+ B_{+}(r-r_c)^{1/2-i\alpha_{l,m}} \hbox{,}\\
\Psi^{\rm Fro}_{l,m-}(r)
=A_{-}(r-r_c)^{1/2+i\alpha_{l,m}}
+B_{-}(r-r_c)^{1/2-i\alpha_{l,m}} \hbox{,}
      \end{array}
  \right.
\end{equation}
where $\Psi^{\rm Fro}_{l,m+}(r)$ (resp. $\Psi^{\rm Fro}_{l,m-}(r)$) is available when $ r>r_c$ (resp. $ r<r_c$). \cite{BookerBretherton1967} and \cite{RingotPhD} both explain a way to connect these solutions, considering that the term $(r-r_c)^{1/2+i\alpha_{l,m}}$ can be compared to an upward propagative wave of the form $e^{ikr}$ and respectively that $(r-r_c)^{1/2-i\alpha_{l,m}}$ can be compared to a downward propagative wave of the form $e^{-ikr}$. We obtain the following identification :
\begin{eqnarray}
\Psi^{\rm Fro}_{l,m\pm}=(r-r_c)^{1/2}&&
\left(\underbrace{A_{\pm}(r-r_c)^{+i\alpha_{l,m}}}_{\hbox{upward
      propagating wave}}\right.\\
&+&\left.\underbrace{B_{\pm}(r-r_c)^{-i \alpha_{l,m}}}_{\hbox{downward
  propagating wave}}\right)\hbox{.}
\end{eqnarray}

In order to connect the solutions, let us observe the comportement of $(r-r_c)$ above and bellow the critical level. As $r-r_c$ decreases from positive to negative values, its complex argument changes continuously from 0 to $-\pi$ \citep{RingotPhD}. Mathematicaly, we get :
\begin{eqnarray}
  \label{eq:argrrc}
  \hbox{if }r>r_c \hbox{: }&& \left(r-r_c\right)^{1/2 \pm i\alpha_{l,m}} =
  |r-r_c|^{1/2 \pm i\alpha_{l,m}} \hbox{,}\\
  \hbox{if }r<r_c \hbox{: }&& \left(r-r_c\right)^{1/2 \pm i\alpha_{l,m}} =
  |r-r_c|^{1/2 \pm i\alpha_{l,m}} e^{-i\pi/2} e^{\pm \pi\alpha_{l,m}} \hbox{.}
\end{eqnarray}

It follows that the solutions above and below the critical layer can be written as
\begin{equation}
\label{Eq:solTGS}
  \left\{
      \begin{array}{l}
\Psi^{\rm Fro}_{l,m+}(r) = A |r-r_c|^{1/2+i\alpha_{l,m}} 
+ B|r-r_c|^{1/2-i\alpha_{l,m}} \hbox{,}\\ 
\Psi^{\rm Fro}_{l,m-}(r)
= -iAe^{\alpha_{l,m}\pi}|r-r_c|^{1/2+i\alpha_{l,m}}
-iBe^{-\alpha_{l,m}\pi}|r-r_c|^{1/2-i\alpha_{l,m}} \hbox{.} 
      \end{array}
  \right.
\end{equation}

\begin{figure}[h]
  \centering
  \includegraphics[width=0.48\textwidth]{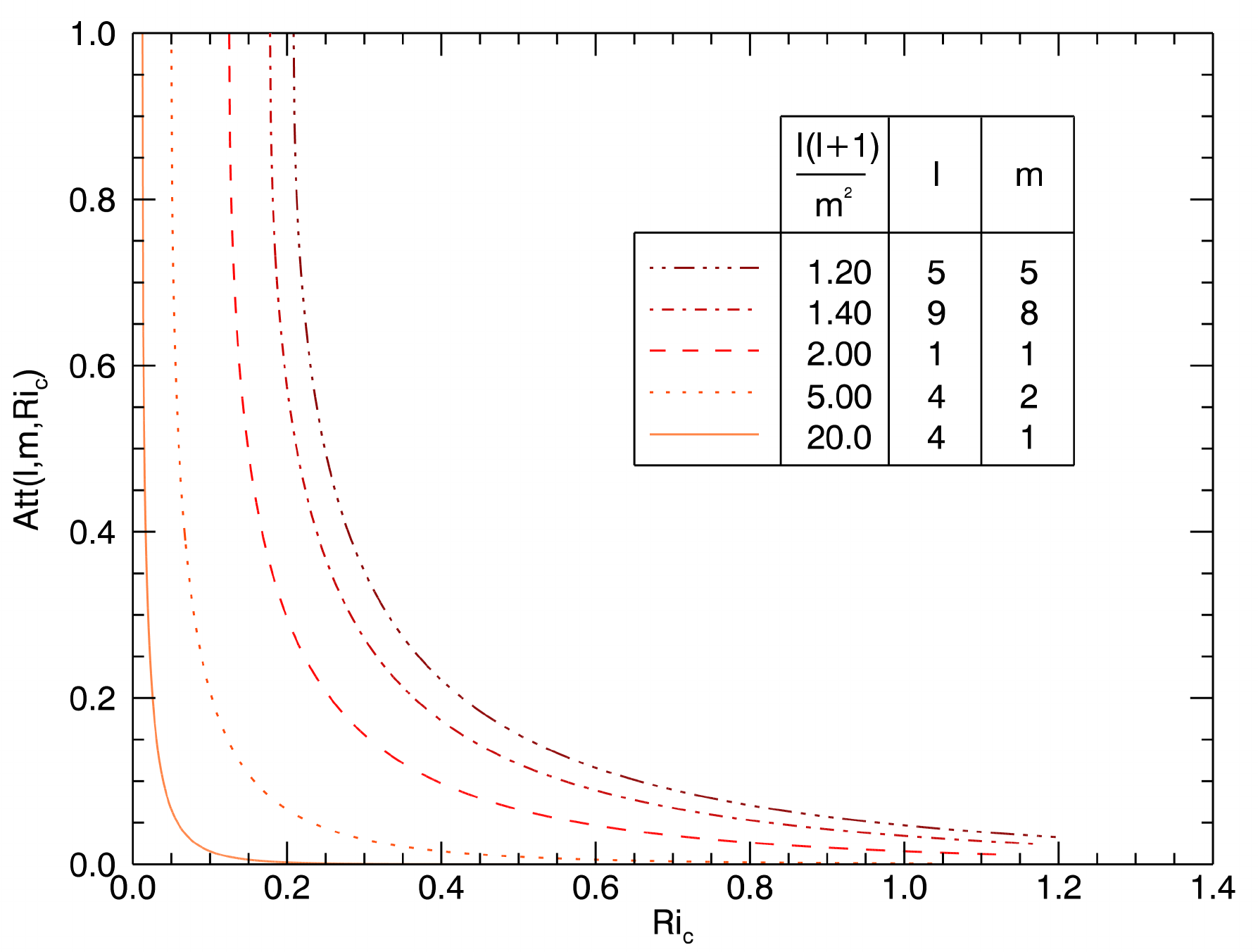}
  \caption{Attenuation rate
    Att=$e^{-\pi\sqrt{\frac{l(l+1)}{m^2}\mathrm{Ri}_c-\frac{1}{4}}}$ of the
    wave passing through a critical layer as a function of the Richardson
    number. We observe that Att increases with
    $\displaystyle\frac{l(l+1)}{m^2}$.}
  \label{fig:1}
\end{figure}

Physically, these equations can be explained this way. Starting from above the critical layer, the downward propagating wave passes through the critical layer and is attenuated
by a factor equals to $e^{-\alpha_{l,m}\pi}$. At the same time, starting from below the critical layer, the upward propagating wave is attenuated by the same factor and its amplitude becomes equal to A. We also underline that both waves take a phase difference when they cross the critical layer. In Fig.~\ref{fig:1}, we represent the attenuation rate of different waves defined by the numbers $l$ and $m$ passing through a critical layer. Greater is the ratio $\frac{l(l+1)}{m^2}$ stronger is the attenuation
Att=$e^{-\pi\sqrt{\frac{l(l+1)}{m^2}\mathrm{Ri}_c-\frac{1}{4}}}$ for the same value of $\mathrm{Ri}_c$. The axis scale depends on $l$ and $m$ because the condition of validity of this result is $\mathrm{Ri}_c >\frac{1}{4}\frac{m^2}{l\left(l+1\right)}$. We so deduce that waves of high ratio $\frac{l(l+1)}{m^2}$ (which not necessary corresponds to high order)
are strongly attenuated, if they reach their critical layer.

We have not yet discussed a latter point: the choice of the method of resolution. Most of the publications concerning IGWs use another process to solve the equation of propagation
\citep{Press1981,ZahnTalonMatias1997,Mathis2009}. In fact, the WKBJ theory
is particularly adapted to the resolution of this equation. However, it is
not convenient in our case because it imposes a condition on the value of
the Richardson number as demonstrated in appendix A. It is shown that the WKBJ approximation is available only if $\mathrm{Ri}_c \gg \frac{1}{4}\frac{m^2}{l(l+1)}$. Despite this restriction, let us write the solution. By separating the domain into two parts, we obtain 
\begin{equation}
\label{eq:solWKBJ}
\Psi^{\rm WKBJ}_{l,m\pm}=\frac{1}{\sqrt{k_V(r)}}(C_{\pm}e^{i\int{k_V(r)dr}}+D_{\pm}e^{-i\int{k_V(r)dr}})\hbox{.}
\end{equation}
As 
\begin{equation}
  \label{eq:2}
  k_V \underset{\mathrm{Ri}_c \gg \frac{1}{4}\frac{m^2}{l(l+1)}}{\longrightarrow} \frac{1}{r-r_c}\sqrt{\frac{l(l+1)}{m^2}\mathrm{Ri}_c} \hbox{,}
\end{equation}
it comes :
\begin{eqnarray}
  \Psi^{\rm WKBJ}_{l,m\pm}=\frac{m}{\sqrt{l(l+1)}\mathrm{Ri}_c}(r-r_c)^{1/2}
\left(C_{\pm}(r-r_c)^{+i\sqrt{\frac{l(l+1)}{m^2}\mathrm{Ri}_c}}\right. \\
\left.+D_{\pm}(r-r_c)^{-i\sqrt{\frac{l(l+1)}{m^2}\mathrm{Ri}_c}}\right)\hbox{.}
\end{eqnarray}

It is comforting to see that both methods (Frobenius and
WKBJ) give the same solution, to a multiplicative constant,
when the Richardson number become high: 
\begin{equation}
\Psi^{\rm Fro}_{l,m}\underset{\mathrm{Ri}_c \gg \frac{1}{4}\frac{m^2}{l(l+1)}}{\longrightarrow} \Psi^{\rm WKBJ}_{l,m} \hbox{.}
\end{equation}

\subsection{The unstable case: $\mathrm{Ri}_c < \frac{1}{4}\frac{m^2}{l(l+1)}$}
In the unstable regime, the Frobenius method also gives a solution but we
are not able to indentify upward and downward propagating waves because of
shear-induced instability and turbulence. As a consequence we can not
connect the solutions at the critical layer. In order to avoid this
difficulty, we here propose to solve Eq. (\ref{Eq:TGS}) following the
method developped by \cite{LindzenBarker1985}. They applied it in the case
of cartesian coordinates and our bringing is to generalize it to spherical
coordinates.  The parameter $\eta_{l,m}$ defined in Eq. \eqref{eq:etalm} is real in this case and we introduce
\begin{equation}
X=k_{Hc}(r-r_c) \hbox{.}
\end{equation}
Equation (\ref{Eq:TGS}) becomes: 
\begin{equation}
\frac{\mathrm d^2 \Psi_{l,m}(X)}{\mathrm dX^2} + \left(\frac{\frac{1}{4}-\eta_{l,m}^2}{X^2}-1 \right) \Psi_{l,m}(X)=0\hbox{.}
\end{equation}

We are seeking solutions of the form $\Psi_{l,m}(X)=X^{\frac{1}{2}}\Phi_{l,m}(X)$, where $\Phi_{l,m}$ is the solution to the Bessel equation
\begin{equation}
X^2\frac{\mathrm d^2}{\mathrm dX^2}\Phi_{l,m}+X\frac{\mathrm d}{\mathrm dX}\Phi_{l,m}-(\eta_{l,m}^2+X^2)\Phi_{l,m}=0\hbox{.}
\end{equation}
Consequently, $\Phi_{l,m}$ is a combination of the Bessel's modified functions $I_{\eta_{l,m}}(X)$ and $I_{-\eta_{l,m}}(X)$: 
\begin{equation}
\Phi_{l,m}=K_1I_{\eta_{l,m}}(X)+K_2I_{-\eta_{l,m}}(X)\hbox{.}
\end{equation}
The final solution is given by:
\begin{equation}
\label{Eq:SolBessel}
\Psi_{l,m}(X)=X^{\frac{1}{2}}\left(K_1I_{\eta_{l,m}}(X)+K_2I_{-\eta_{l,m}}(X)\right)\hbox{.}
\end{equation}

\begin{figure}[h]
  \centering
  \includegraphics[width=0.5\textwidth]{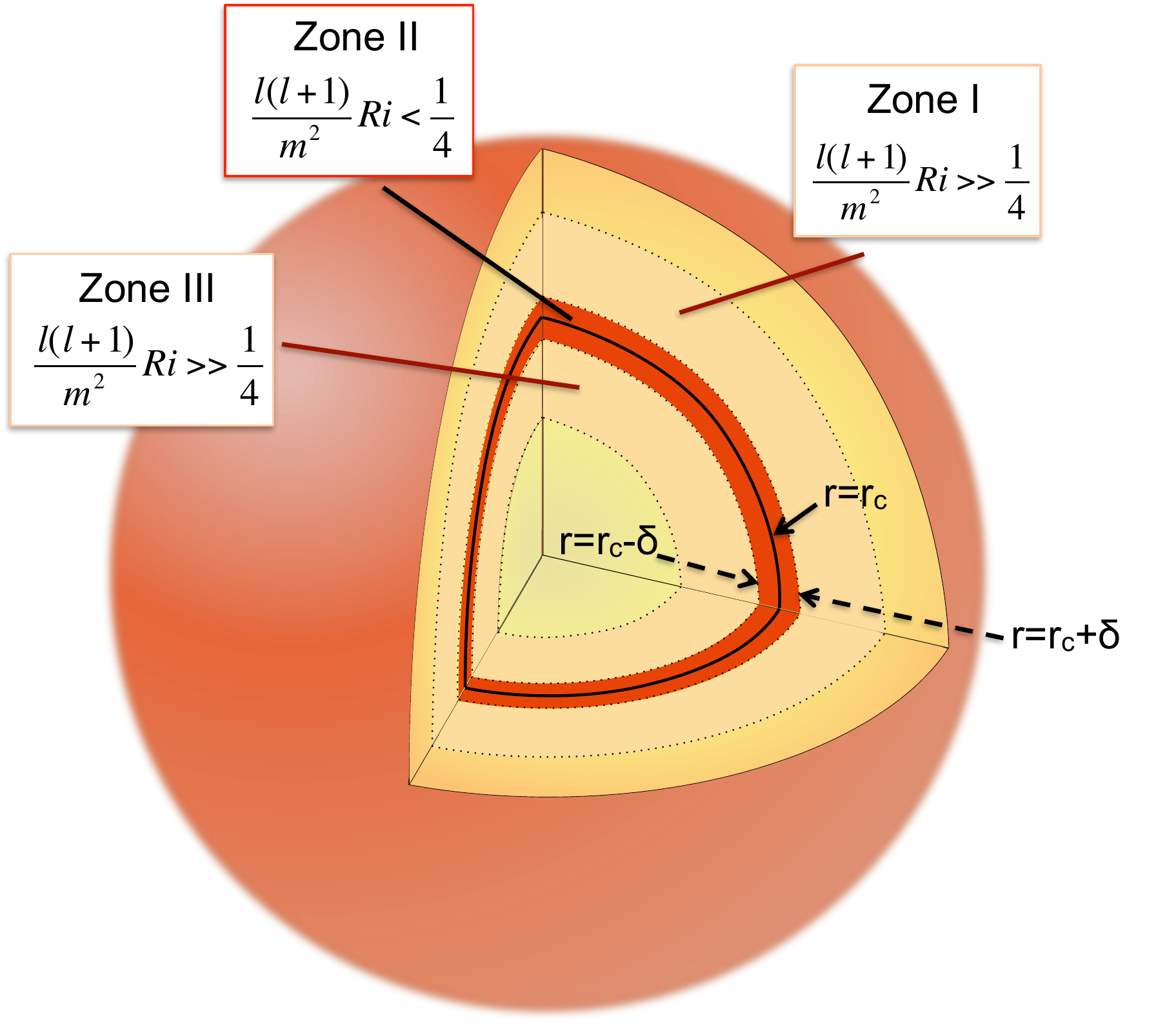}
  \caption{Assumed neighbourhood of an unstable critical layer for the
    calculation of the IGWs' reflection and transmission coefficients. We assume that the unstable region around the critical layer (in red) has a thickness given by $2\delta$ (Zone II). The surrounding regions where IGWs are propagative are in beige (Zones I and III).}
  \label{fig:3}
\end{figure}

We would like to calculate the reflection and transmission coefficients of
the wave passing through the unstable critical layer. We assume that the
fluid has the profile described in Fig.~\ref{fig:3}. We decompose the studied region into three zones defined by the value of
the quantity $\frac{l(l+1)}{m^2}\mathrm{Ri}_c$. In zones I and III, the
Richardson number is high enough to allow us to apply the WKBJ method described in the previous part. In unstable zone II around the studied critical layer however, we must use the solution with the modified Bessel functions. Moreover, we assume here that the thickness $2\delta$ of this latter is small in comparison with the characteristic length of the problem. Then, we can consider that the wavenumber $k_V$ is constant as
\begin{equation}
\label{Eq:kvc}
k_{Vc}^2=\frac{l(l+1)}{m^2}\frac{\mathrm{Ri}_c}{\delta^2}-\frac{l(l+1)}{r_c^2}\hbox{.}
\end{equation}
Let us consider a wave coming from the overside of the critical layer (zone I). It is partly transmitted toward zone III and partly reflected backward zone I.  We so write the solutions corresponding to the three zones:  
\begin{equation}
  \left\{
      \begin{array}{l}
\Psi_{I}(r)=e^{-i k_V (r-r_c)}+Re^{i k_V (r-r_c)} \hbox{,}\\
\Psi_{II}(r)=(r-r_c)^{1/2}\left[AI_{\eta_{l,m}}(k_{Hc}(r-r_c))+BI_{-\eta_{l,m}}(k_{Hc}(r-r_c))\right] \hbox{,} \\
\Psi_{III}(r)=Te^{i  k_V (r-r_c)} \hbox{,}
      \end{array}
  \right.
\end{equation}
where $\Psi_{I}(r)$ is available if $(r-r_c)>\delta$, $\Psi_{II}(r)$ if $-\delta<(r-r_c)<\delta$ and $\Psi_{III}(r)$ if $(r-r_c)>-\delta$. The coefficients $A$, $B$, $T$ and $R$ are calculated thanks to the {following} four continuity equations:
\begin{equation}
  \left\{
      \begin{array}{l}
\Psi_I(r_c+\delta)=\Psi_{II}(r_c+\delta)\hbox{,} \\
\displaystyle \frac{\mathrm d \Psi_I}{\mathrm dr}(r_c+\delta) = \displaystyle \frac{\mathrm d \Psi_{II}}{\mathrm dr}(r_c+\delta)\hbox{,} \\
\Psi_{III}(r_c-\delta)=\Psi_{II}(r_c-\delta)\hbox{,} \\
\displaystyle \frac{\mathrm d \Psi_{III}}{\mathrm dr}(r_c-\delta) = \displaystyle \frac{\mathrm d \Psi_{II}}{\mathrm dr}(r_c-\delta)\hbox{.}
      \end{array}
  \right.
\end{equation}

\begin{figure*}
\begin{center}
   \includegraphics[width=0.435\textwidth]{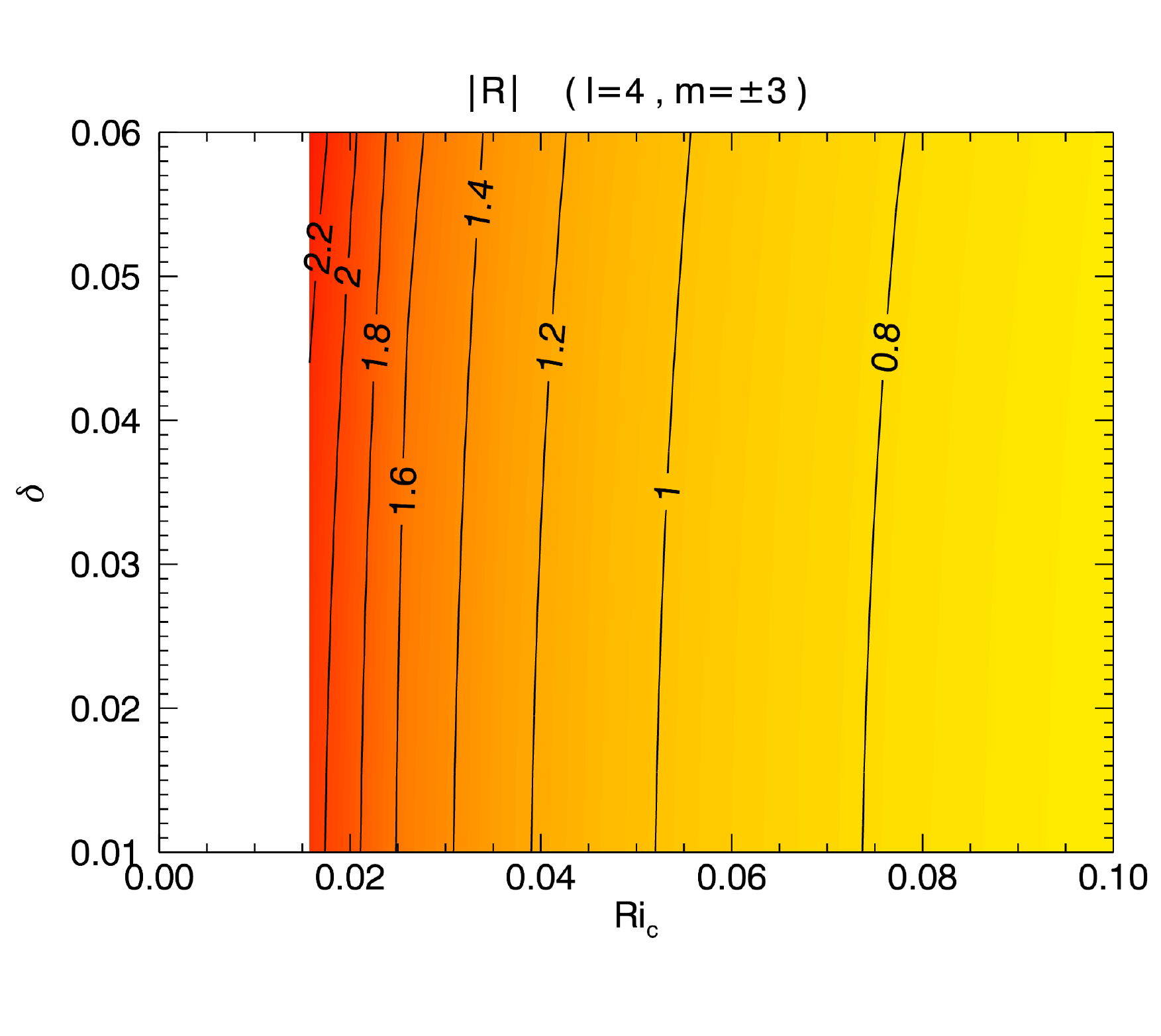}
   \includegraphics[width=0.475\textwidth]{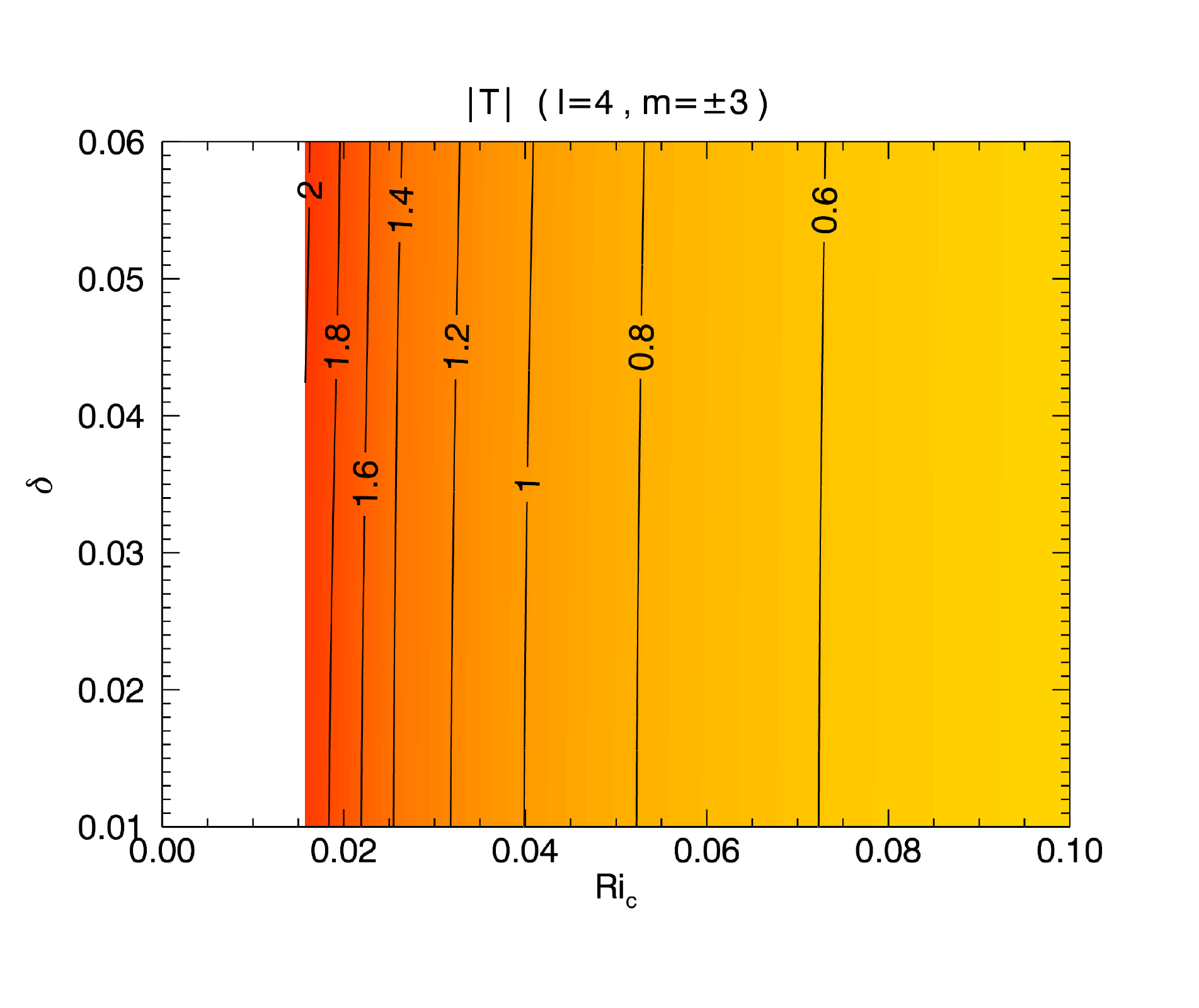} \\
   \includegraphics[width=0.45\textwidth]{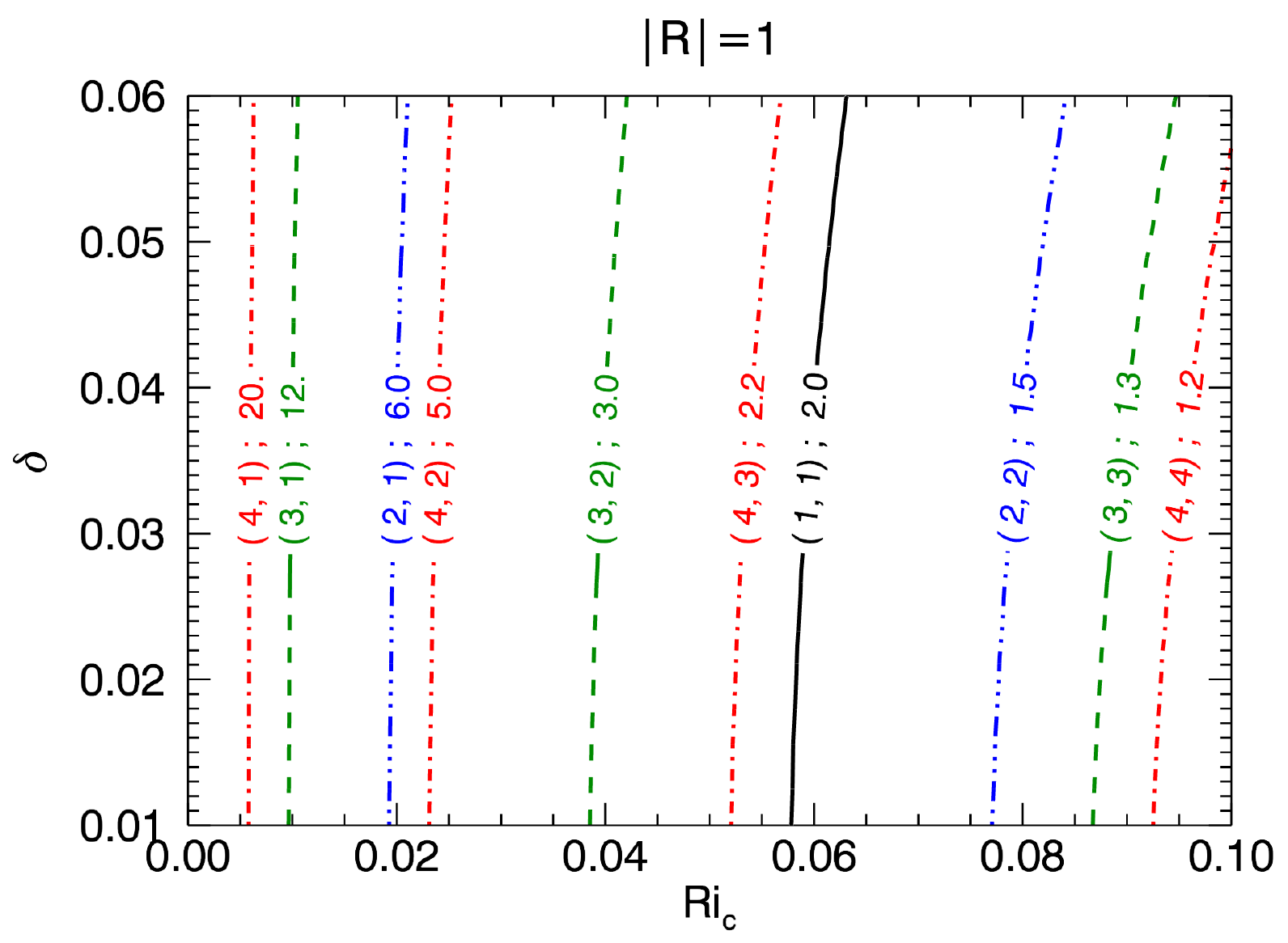}
   \includegraphics[width=0.45\textwidth]{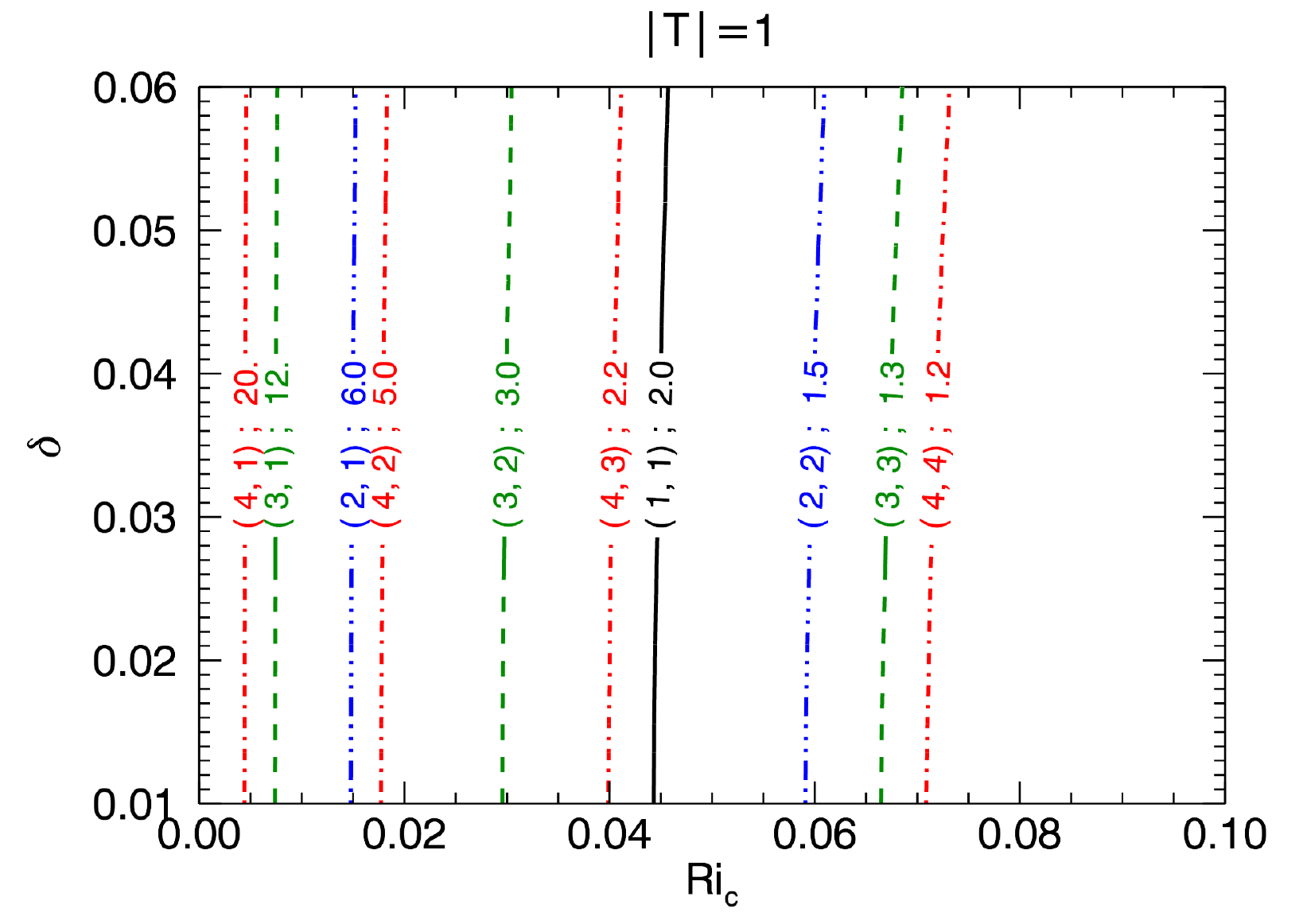}
   \caption{Level lines of reflection $|R|$ and transmission $|T|$
     coefficients of an IGW at a critical layer as a function of the
     Richardson number $\mathrm{Ri}_c$ and of the thickness $\delta$ of the critical
     layer (zone II in Fig.~\ref{fig:3}). The top panels represent the
     level lines for an arbitrary choosen value of $(l,m)=(4,\pm 3)$, while
     the bottom panels show levels lines $|R|=1$ and $|T|=1$ for different
     couples $(l,m)$.}
   \label{fig:4}
\end{center} 
\end{figure*}

They correspond to the continuity of the solution and of its first derivative, which physically means that both displacement and mechanical {stresses} are continuous. After some algebra, we obtain the coefficients $R$ and $T$ of reflection and transmission, depending on the stiffness $\delta$, on the vertical and horizontal wavenumbers $k_{Vc}$ and
$k_{Hc}$ (see Eq. (\ref{Eq:kvc}) and  (\ref{Eq:khc})) , and on the variable $\eta_{l,m}$ (Eq. (\ref{eq:etalm})). In order to lighten the formula, we note $I_{\pm\eta_{l,m}}$ instead of
$I_{\pm\eta_{l,m}}(k_{Hc}\delta)$. {Then, we obtain}  
\begin{equation}
     R=\frac{R_{\mathrm{num}}}{R_{\mathrm{denom1}}+R_{\mathrm{denom2}}} \hbox{,}
\end{equation}
with 
\begin{eqnarray*}
&&R_{\mathrm{num}}\!=\!\left[k_{Hc}I_{\eta_{l,m}}'+\left(\frac{1}{2\delta}-i
  k_{Vc}\right)I_{\eta_{l,m}}\right]\!\left[k_{Hc}I_{-\eta_{l,m}}'+\left(\frac{1}{2\delta}-ik_{Vc}\right)I_{-\eta_{l,m}}\right]\!\hbox{,}\\
&&R_{\mathrm{denom1}}=k_{Hc}^2I_{-\eta_{l,m}}'I_{-\eta_{l,m}}+\left(\frac{1}{4\delta^2}+k_{Vc}^2\right)I_{-\eta_{l,m}}I_{\eta_{l,m}}\hbox{,}\\
&&R_{\mathrm{denom2}}=-\frac{2k_{Vc}}{\pi\delta}\cos\left(\eta_{l,m}\pi\right)+\frac{k_{Hc}}{2\delta}\left(I_{\eta_{l,m}}I_{-\eta_{l,m}}'+I_{-\eta_{l,m}}I_{\eta_{l,m}}'\right)
\end{eqnarray*}
and
\begin{equation}
      T=\frac{T_{\mathrm{num}}}{T_{\mathrm{denom}}}\hbox{,}
\end{equation}
with 
\begin{eqnarray*}
&&T_{\mathrm{num}}=\frac{2i k_{Vc}}{\delta \pi} \hbox{,} \\
&&T_{\mathrm{denom}}=R_{\mathrm{denom1}}+R_{\mathrm{denom2}} \hbox{.}
\end{eqnarray*}
We have now calculated the transmission and reflection coefficients of an IGW, under some assumption, {through an unstable region around a given critical layer} where
$\mathrm{Ri}_c<\frac{1}{4}\frac{m^2}{l(l+1)}$. We represent the level lines of $|R|$ and $|T|$ in Fig.~\ref{fig:4} for $l=4$ and $m=\pm3$. They are plotted as function of the Richardson number at the critical layer $\mathrm{Ri}_c$, growing from 0 to its maximum value defined by $\frac{l(l+1)}{m^2}{\mathrm{Ri}_c}_{max}=\frac{1}{4}$. For instance, in Fig.~\ref{fig:4}, ${\mathrm{Ri}_{c}}_{max}=\frac{1}{4}\frac{3^2}{4(4+1)}\approx 0.11$. The other variable is the half-thickness $\delta$ of the {unstable} layer (zone II), arbitrarily
choosen, {that points the non-local character of unstable turbulent
  layers.} Let us underline the main result: both coefficients are greater
than 1 when $\frac{l(l+1)}{m^2}\mathrm{Ri}_c$ is small
enough. Consequently, for a low Richardson number at the critical layer,
the wave can be over-reflected and over-transmitted at the same time. It
means that on the contrary of the first stable case, the wave take
{potential energy from the unstable fluid and convert it into kinetic
  energy. In other words, the turbulent layer acts as an excitation
  region.} If $|R|<1$ and $|T|<1$, we speak about IGWs  "tunneling" \citep{SutherlanYewchuk2004,BrownSutherland2006,NaultSutherland2007}. Another remark concerns de dependency of $|R|$ and  $|T|$ with $m$. We denote that the sign of $m$ does not matter since only its square appears in the expressions. Physically, it shows that the critical layer's action is the same on prograde and retrograde waves. This point is of importance because we know that other
dissipative processes occuring during the propagation of IGW discriminate between both types of waves. 
In order to visualize the action of the critical layer on differents waves, Fig.~\ref{fig:3} shows the level lines $|R|=1$ for $ 1 \leq l \leq 5 $ and  $ 1 \leq m \leq l$. We previously said that it is useless to consider negative values of $m$ since $|R|$ depends only on $m^2$. The pair ($l$,$m$) is indicated on each ligne, followed by the value of $\frac{l(l+1)}{m^2}$. Lines in the same color correspond to the same value of $l$. These lines mark out the limit to observe over-reflection, for a chosen wave. We observe that higher is the value of $\frac{l(l+1)}{m^2}$, stronger is the condition on $\mathrm{Ri}_c$ to observe an over-reflection.

\subsection{Choice of the method}
We have decided to apply different methods to solve the stable and unstable cases. However, it could be legitimate to wonder if both methods are equivalent from a mathematical point of view. In this part, we present a short comparison between the solutions obtained with the method of Frobenius and the one with Bessel functions. The modified Bessel
function $I_{\eta_{l,m}}(X)$ can be computed using
\begin{equation}
  I_{\eta_{l,m}}(X) = \left(\frac{1}{2}X\right)^{\eta_{l,m}}\sum\limits_{k=0}^{\infty} 
\frac{\left(\frac{1}{2}X\right)^{2k}}{k!\Gamma(\eta_{l,m}+k+1)} \hbox{.}
\end{equation}
At the neighbourhood of the critical layer, X tends to $0$ and the first-order expression is
\begin{equation}
  I_{\eta_{l,m}}(X) = \left(\frac{1}{2}X\right)^{\eta_{l,m}}
\left(\frac{1}{\Gamma(\eta_{l,m}+1)} 
+ O(X^2)\right)
\hbox{.}
\end{equation}
Close to the critical layer, the global solution given in Eq. (\ref{Eq:SolBessel}) is then  
\begin{equation}
\Psi_{l,m}^{\rm Bessel}(X)=
\left(
\Delta_1 \frac{X^{\frac{1}{2}+\eta_{l,m}}}{\Gamma(\eta_{l,m}+1)} 
+\Delta_2 \frac{X^{\frac{1}{2}-\eta_{l,m}}}{\Gamma(1-\eta_{l,m})}
+ O(X^2)\right)\hbox{.}
\end{equation}
Now, let us remind the expression of the solution given by the method of Frobenius (Eq. (\ref{Eq:case2})), rewritten with the previous notations
\begin{equation*}
\Psi_{l,m}^{\rm Fro}(r) = A_2 |r-r_c|^{1/2+\eta_{l,m}} + B_2 |r-r_c|^{1/2-\eta_{l,m}}.
\end{equation*}
In conclusion, for a fixed couple $(l,m)$, $\Psi_{l,m}^{\rm Fro}(r)$ and $\Psi_{l,m}^{\rm Bessel}(X)$ vary in the same way as function of $r-r_c$.

\section{Case of the non-perfect fluid}
From now, we have studied the role of the critical layers assuming that the fluid was perfect. In order to make the problem more realistic we include in the second part the viscosity $\nu$ of the fluid and the coefficient of thermal conductivity $\kappa$ \citep[e.g.][]{Koppel1964,Hazel1967,BaldwinRoberts1970,VanDuinKelder1986}.

\subsection{Equation of propagation of IGW near a critical layer}
The linearized equations of hydrodynamics in Eq. (\ref{Eq:HydroP}) become: 
\begin{equation}
  \left\{
      \begin{array}{l}
D_t\vec{u}=-\frac{\vec{\nabla}p'}{\bar\rho}+\frac{\rho'}{\bar\rho}\vec{g}+\nu\Delta\vec{u}\hbox{,}\\ 
\label{Eq:QtyMvt}
D_t\rho'+\vec{\nabla}.(\bar\rho\vec{u})=0\hbox{,}\\ 
D_t\left(\frac{\rho'}{\bar\rho}-\frac{1}{\Gamma_1}\frac{p'}{\bar p}\right)-\frac{N^2}{g}u_r=\frac{\kappa}{\bar\rho}\Delta\rho'\hbox{.}
      \end{array}
    \right.
\end{equation}
The following method for the building of the propagation equation is
adapted from the work of \cite{BaldwinRoberts1970}. We assume that the
mean density $\bar\rho$ of the fluid nearly takes a constant value, that is to say
that $\frac{1}{\bar\rho}\frac{\mathrm{d}\bar\rho}{\mathrm{d}r}$ is small
compared with others characteristic lenghts. 
First, we project
Eq. (\ref{Eq:QtyMvt}) onto $\hat{e}_r$  and apply the operator
$\vec\nabla^2$.
Then we apply $D_t-\kappa \vec\nabla^2$. After combination with the two other equations, we obtain :
\begin{equation}
\label{eq:TGSint}
(D_t-\kappa \vec\nabla^2)\left[(D_t-\nu \vec\nabla^2)(\vec\nabla^2 u_r +
  (\Omega''+\frac{2}{r}\Omega')\partial_{\varphi} u_r\right] = N^2 \vec\nabla_\perp^2 u_r \hbox{.}
\end{equation}
As done in the previous part, 
we decompose the radial velocity on the basis of spherical harmonics 
\begin{eqnarray}
\label{Eq:53}
u_r(r,\theta,\varphi,t) &=& \sum\limits_{l,m}\hat{u}_{r;l,m}(r)Y_{l,m}(\theta,\varphi)e^{i\sigma_{w}
     t}\hbox{,}
\end{eqnarray}
where $\sigma=\sigma_{w}+m\Delta\Omega(r)$.
Moreover, it is easier to work with dimensionless numbers. For this reason, we introduce the notations detailed in Tab. \ref{tab:notations}. 
\begin{table}[h]
\centering
\caption{Dimensionless numbers used for the resolution of Eq. (\ref{eq:TGSint}). $L$ and $V$ are respectively the length and velocity scales.}
\begin{tabular}[h]{|c|c|c|}\hline
$P_{\mathrm r}$&$\nu/\kappa$ & Prandtl number\\\hline
$R_{\mathrm e}$&$VL/\nu$ & Reynolds number \\\hline
$\mathrm{Ri}$&$\left(LN/V\right)^2$ & Richardson number \\\hline
\end{tabular}
\label{tab:notations}
\end{table}

Then, Eq. \ref{eq:TGSint} becomes for each pair ($l$,$m$) such as $l\in\mathbb{N}$ and $m \in \llbracket
-l,l \rrbracket$ : 
\begin{eqnarray}
&&\left(\Delta_{l}-k_H^2-i\sigma R_{\mathrm e}P_{\mathrm r}\right)
\left(\Delta_{l}-k_H^2-i\sigma R_{\mathrm e}\right)
\left(\Delta_{l}-k_H^2\right)\hat{u}_{r;l,m}\nonumber \\
&&= -k_H^2R_{\mathrm e}^2P_{\mathrm r} N^2\hat{u}_{r;l,m} \hbox{,}
\end{eqnarray}
where $\Delta_l$ is the scalar spherical Laplacian operator :
\begin{equation}
  \label{eq:1}
  \Delta_l = \partial_{rr}^2 + \frac{2}{r}\partial_r - \frac{l(l+1)}{r^2}\hbox{.}
\end{equation}

Lastly, we introduce $\eta=(im\Omega_c' R_{\mathrm e})^{1/3}(r-r_c)$ and thanks to a developpement close to the critical layer we obtain the sixth-order equation: 
\begin{eqnarray}
\label{Eq:TGSNP}
&&\frac{1}{P_{\mathrm r}}\frac{\partial^6\chi_{l,m}}{\partial\eta^6}
-\eta\left(1+\frac{1}{P_{\mathrm r}}\right)\frac{\partial^4\chi_{l,m}}{\partial\eta^4}
-\frac{2}{P_{\mathrm r}}\frac{\partial^3\chi_{l,m}}{\partial\eta^3}\nonumber\\
&+&\eta^2\frac{\partial^2\chi_{l,m}}{\partial\eta^2}
+\frac{l(l+1)}{m^2}\mathrm{Ri}_c\chi_{l,m}
=0\hbox{,}
\end{eqnarray}
where {now} $\chi_{l,m}=\bar\rho^{1/2}\eta\hat{u}_{\eta;l,m}$. Equation (\ref{Eq:TGSNP}) can be compared with the one obtained by \cite{Press1981} who has neglected the viscosity $\nu$. Moreover, we can see that if we take $\nu=\kappa=0$, and without forgetting that $P_r$ depends on $\nu$ too, Eq. (\ref{Eq:TGSNP}) is identical to Eq. (\ref{Eq:TGS}) for the perfect fluid.

\subsection{Mathematical resolution}
The resolution of Eq. (\ref{Eq:TGSNP}) requests several substitutions and is quite complex. The detailed calculation can be found in appendix B and we give here only the main steps. We draw our inspiration from \cite{Hazel1967,BaldwinRoberts1970,Koppel1964,VanDuinKelder1986} who solve the same equation in cartesian coordinates. {However,} the resolution in spherical coordinates has some differences. The aim is {here} to rewrite the equation under a known form: the Whittaker differential equation; {then,} after some algebra, we can show that Eq. (\ref{Eq:TGSNP}) can be written in the following form
\begin{equation}
\label{Eq:Whittaker2}
\frac{\mathrm d^2V}{\mathrm ds^2}
+\left(\frac{\frac{1}{4}-M_{l,m}^2}{s^2}+\frac{\Lambda}{s}-\frac{1}{4}\right)V=0 \hbox{.}
\end{equation} 
The solutions of Eq. (\ref{Eq:Whittaker2}) are {thus} the Whittaker functions \citep{HandbookMathFunctions}:
\begin{equation}
V_{\Lambda,M_{l,m}}(s)=e^{-\frac{s}{2}}s^{\frac{1}{2}+M_{l,m}}{_{1}F_{1}}\left(\frac{1}{2}+M_{l,m}+\Lambda;1+2M_{l,m};s\right) \hbox{,}
\end{equation}
where $_{1}F_{1}$ is the confluent hypergeometric function of Kummer
\begin{equation}
_{1}F_{1}(a;b;z)= \sum \limits_{n=0}^{\infty}\frac{(a)_n}{(b)_n}\frac{z^n}{n!}
\end{equation}
with $(k)_n=\prod\limits_{i=0}^{n-1}(k+i)=\frac{\Gamma\left(k+n\right)}{\Gamma\left(k\right)}$ and $(k)_0=1$, $\Gamma\left(z\right)$ being the usual Gamma function,
\begin{equation}
M_{l,m}^2=\frac{1}{4}-\frac{2+\frac{l(l+1)}{m^2}\mathrm{Ri}_c}{9}\hbox{,}
\end{equation}
and
\begin{equation}
\Lambda=-\frac{1}{3}\hbox{.}
\end{equation}

Thanks to this solution, we obtain the expression of the radial displacement of the wave: 
\begin{eqnarray}
\bar{\rho}^{1/2}\eta^2\lefteqn{\hat{\xi}_{\eta;l,m}\!\propto\!}\nonumber\\
f(\kappa,P_{\rm r})&\int_{a}^{b}&{e^{\eta t-t^3/3}t^{3M_{l,m}-3/2}{_{1}F_{1}}\left(\frac{1}{6}+M_{l,m};1+2M_{l,m};t\right)dt}\,\hbox{;}\nonumber\\
\end{eqnarray}
we will clarify the function $f(\kappa,P_{\rm r})$ of the thermal diffusivity coefficient later.

There is still the last step to get over. We must find a curve along which we integrate the solution, that is to say, we determine $a$ and $b$. During the calculation detailed in the appendix B, Eq. \eqref{Eq:Conditionab}: 
\begin{eqnarray}
\left[-\left(1+\frac{1}{P_{\mathrm r}}\right)t^4ve^{\eta t}-\frac{\mathrm d}{\mathrm dt}\left(t^2v\right)e^{\eta t}+zt^2ve^{\eta t}\right]_a^b=0
\end{eqnarray}
has not been used yet. A sufficient condition to make this egality true is
$\frac{-t^3}{3}\underset{|t|\to +\infty}{\longrightarrow} \infty$. 
Considering that $t^3=\vert t \vert^3e^{3i\theta_t}$ ($\theta_t$ being the complex argument of $t$), we have
\begin{equation}
\frac{-t^3}{3}\underset{|t|\to +\infty}{\longrightarrow}\infty \hbox{  }
\Leftrightarrow \hbox{  } \theta_t \equiv 0\left[\frac{2\pi}{3}\right]\hbox{.}
\end{equation}
There are several possibilities for the choice of the curves. \cite{Koppel1964} propose to build a basis of six solutions using the curves  ($\mathcal{C}_1$,$\mathcal{C}_2$,$\mathcal{C}_3$) represented in Fig.~\ref{fig:6} (left). Therefore, the solutions of the sixth order equation (Eq. (\ref{Eq:TGSNP})) are linear combinations of 
\begin{eqnarray}
\label{Ui}
\lefteqn{U_i(\eta) = f(\kappa,P_{\rm r})}\nonumber\\
& &\times\int_{\mathcal{C}_i}{e^{\eta
    t-t^3/3}t^{3M_{l,m}-3/2}{_{1}F_{1}}\left(\frac{1}{6}M_{l,m};1+2M_{l,m};t\right)}\,\mathrm
d t
\end{eqnarray}
and $V_i(\eta)$ with $i \in \{1,2,3\}$, where $V_i(\eta)$ corresponds to $U_i(\eta)$ with the opposite sign for $M_{l,m}$. 

\begin{figure}
\begin{center}
   \includegraphics[width=0.24\textwidth]{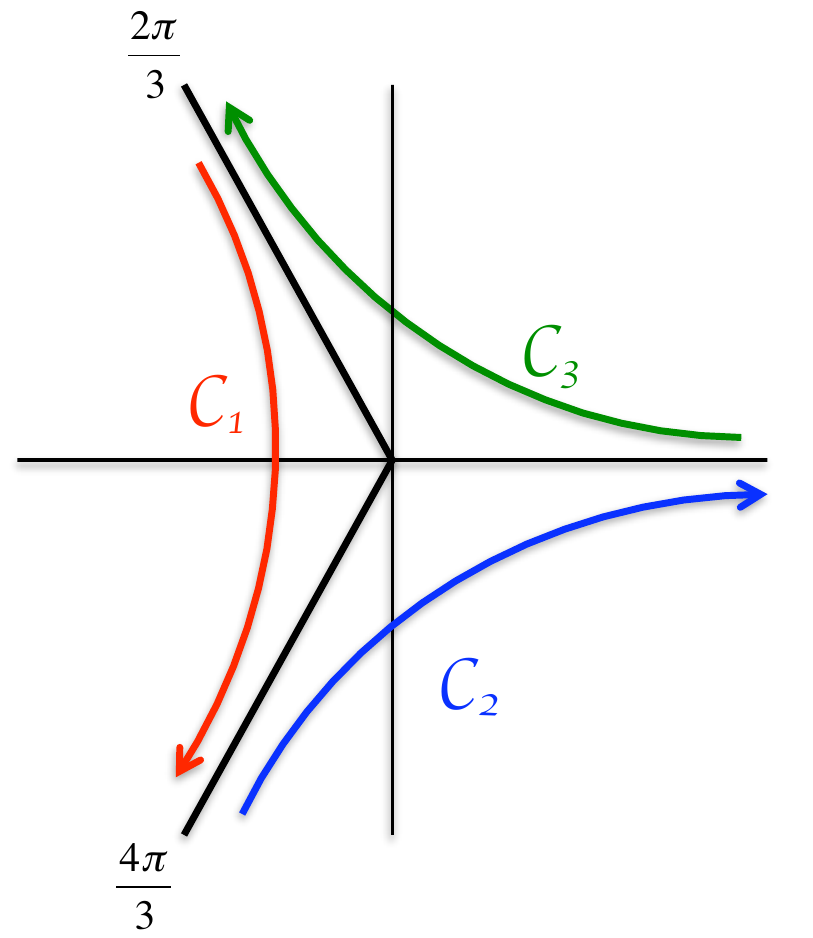}
   \includegraphics[width=0.24\textwidth]{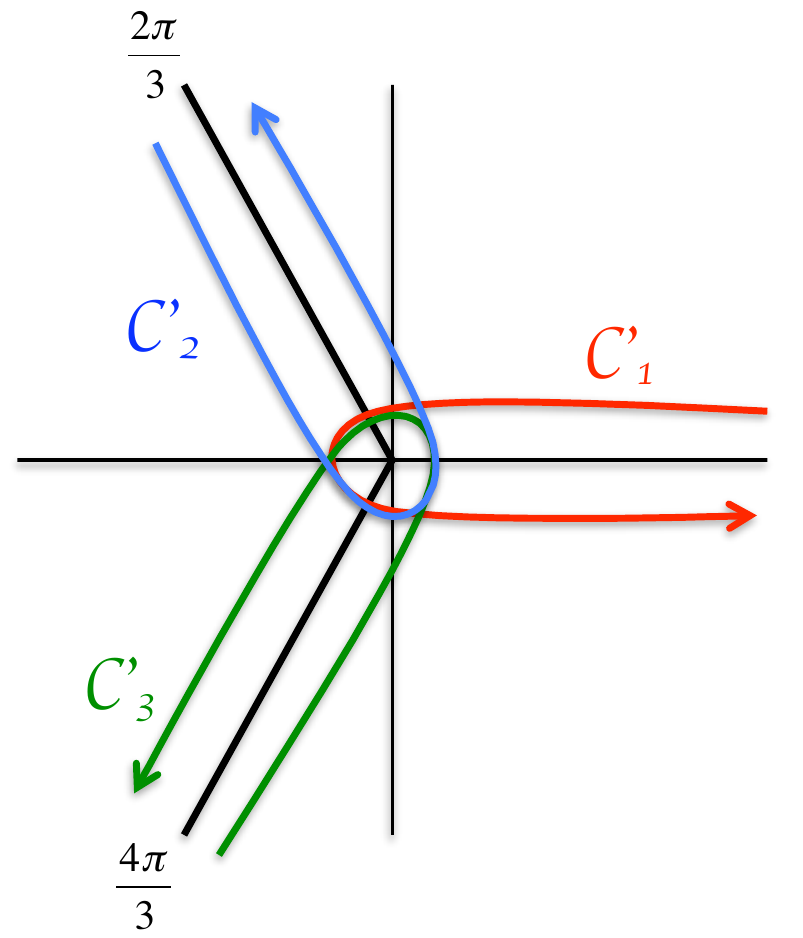}
   \caption{Left: curves ($\mathcal{C}_1$,$\mathcal{C}_2$,$\mathcal{C}_3$)
     defining the basis ($U_1,U_2,U_3,V_1,V_2,V_3$). Right: curves
     ($\mathcal{C}'_1$,$\mathcal{C}'_2$,$\mathcal{C}'_3$) defining the
     basis ($u_1,u_2,u_3,v_1,v_2,v_3$).} 
    \label{fig:6}
\end{center} 
\end{figure} 

\subsection{Application to IGWs}
It is now time to apply these mathematical results to {IGWs}. With this aim in view, let us remind the solution for the perfect fluid obtained in the first part. For the moment, it is not necessary to distinguish between {the stable regime, i.e.} $\frac{l(l+1)}{m^2}\mathrm{Ri}_c > \frac{1}{4}$, and {the unstable one, i.e.} $\frac{l(l+1)}{m^2}\mathrm{Ri}_c < \frac{1}{4}$. The Frobenius solution is available everywhere if we take $\eta_{l,m}=\sqrt{\frac{1}{4}-\frac{l(l+1)}{m^2}\mathrm{Ri}_c}$ as a complex number. Remembering that
$\Psi_{l,m}(r)=\bar\rho^{\frac{1}{2}}r^2\hat\xi_{r;l,m}(r)$, the radial
Lagrangian displacement is
\begin{equation}
  \left\{
      \begin{array}{l}
\hat\xi_{\mathrm P+} =\frac{1}{\bar\rho^{1/2}r^2} \left(
A |r-r_c|^{1/2+i\alpha_{l,m}} 
+ B|r-r_c|^{1/2-i\alpha_{l,m}}
\right)\\
\hat\xi_{\mathrm P-}= 
\frac{1}{\bar\rho^{1/2}r^2}\left(
-iAe^{\alpha_{l,m}\pi}|r-r_c|^{1/2+i\alpha_{l,m}}
-iBe^{-\alpha_{l,m}\pi}|r-r_c|^{1/2-i\alpha_{l,m}}
\right)\hbox{.}
      \end{array}
  \right.
\end{equation}
In order to make the reading easier, we have removed the indices $r;l,m$. Indice $P$ designates the solution for the perfect fluid.\\

Concerning the solution for non-perfect fluids, \cite{Hazel1967} proposes to use another basis of the solutions of Eq. (\ref{Eq:TGSNP}). The curves ($\mathcal{C}'_1$,$\mathcal{C}'_2$,$\mathcal{C}'_3$) are represented in Fig.~\ref{fig:6}. The new basis  ($u_1,u_2,u_3,v_1,v_2,v_3$) is defined by: 
\begin{itemize}
   \item $u_1=U_1+U_2+U_3$ and $v_1=V_1+V_2+V_3$,
   \item  $u_2=U_1+U_2+U^*_3$ and $v_2=V_1+V_2+V^*_3$,
   \item  $u_3=U_1+U^*_2+U_3$ and $v_3=V_1+V^*_2+V_3$,
\end{itemize}
where $X^*$ is the complex conjugate of $X$. Consequently, the solution for the non-perfect fluid (indice NP) can be written as :
\begin{equation}
\hat\xi_{\mathrm{NP}\pm}=\frac{1}{\bar\rho^{1/2}\eta^2}\sum \limits_{i=1}^{3}\left[\alpha_{i\pm}u_i\left(\eta\right)+\beta_{i\pm}v_i\left(\eta\right)\right]\hbox{.}
\end{equation}
A relation between both solutions $\hat\xi_{\mathrm P\pm}$ and $\hat\xi_{\mathrm{NP}\pm}$ exists if we consider that the Reynolds number is great. {In the case of the Sun \citep[e.g.][]{BrunZahn2006}}, the {microscopic} viscosity in the radiative zone is weak. For this reason, it is appropriate to consider that the Reynolds number $R_{\mathrm e}=\frac{VL}{\nu}$ is much greater than 1. This assuption leads to the relation
\begin{equation}
\hat\xi_{\mathrm{NP}\pm}\underset{R_{\mathrm e}\to +\infty}{\longrightarrow}\hat\xi_{\mathrm P\pm}\hbox{.}
\end{equation}
\cite{BaldwinRoberts1970} give tables for the asymptotic comportement of $u_i$ and $v_i$ ( i$\in{1,2,3}$). The solutions $u_1$, $u_2$, $v_1$ et $v_2$ diverge when $R_{\mathrm e}\to +\infty$. They are {therefore} physically unacceptable and we deduce {that} 
\begin{eqnarray}
\hat\xi_{\mathrm{NP}\pm}=\frac{1}{\bar\rho^{1/2}\eta^2}\alpha_{3\pm}u_3(\eta)+\beta_{3\pm}v_3(\eta)\hbox{.}
\end{eqnarray}
Tables (\ref{tab:asymp1}) and (\ref{tab:asymp2}) give the asymptotic expressions of $u_3$ and $v_3$ as a function of the sign of  $r-r_c$.
\begin{table}[h]
\centering
\caption{Expressions of  $u_3$ and $v_3$ when $R_{\mathrm e}\to +\infty$, below the critical layer.}
\begin{tabular}{|c|c|}\hline
\multicolumn{2}{|c|}{$r>r_c$}\\ \hline
&\\
$u_3(\eta)$&$\displaystyle\frac{2i\pi}{\Gamma(\frac{3}{2}-i\alpha_{l,m})}(im\Omega_c'R_{\mathrm e})^{\frac{1}{3}\left(\frac{1}{2}-i\alpha_{l,m}\right)}|r-r_c|^{\frac{1}{2}-i\alpha_{l,m}}$\\ 
&\\\hline
&\\
$v_3(\eta)$&$\displaystyle\frac{2i\pi}{\Gamma(\frac{3}{2}+i\alpha_{l,m})}(im\Omega_c'R_{\mathrm e})^{\frac{1}{3}\left(\frac{1}{2}+i\alpha_{l,m}\right)}|r-r_c|^{\frac{1}{2}+i\alpha_{l,m}}$\\ 
&\\\hline
\end{tabular}
\label{tab:asymp1}\\

\caption{Expressions of  $u_3$ and $v_3$ when $R_{\mathrm e}\to +\infty$, above the critical layer.}
\begin{tabular}{|c|c|}\hline
\multicolumn{2}{|c|}{$r<r_c$}\\ \hline
&\\
$u_3(\eta)$&$\displaystyle\frac{2\pi}{\Gamma(\frac{3}{2}-i\alpha_{l,m})}e^{-\alpha_{l,m}\pi}(im\Omega_c'R_{\mathrm e})^{\frac{1}{3}\left(\frac{1}{2}-i\alpha_{l,m}\right)}|r-r_c|^{\frac{1}{2}-i\alpha_{l,m}}$\\ 
&\\\hline
&\\
$v_3(\eta)$&$\displaystyle\frac{2\pi}{\Gamma(\frac{3}{2}+i\alpha_{l,m})}e^{-\alpha_{l,m}\pi}(im\Omega_c'R_{\mathrm e})^{\frac{1}{3}\left(\frac{1}{2}+i\alpha_{l,m}\right)}|r-r_c|^{\frac{1}{2}+i\alpha_{l,m}}$\\ 
&\\\hline
\end{tabular}
\label{tab:asymp2}

\end{table}\\

In the {stable} case where $\frac{l(l+1)}{m^2}\mathrm{Ri}_c\geq\frac{1}{4}$, $\alpha_{l,m}=\sqrt{\frac{l(l+1)}{m^2}\mathrm{Ri}_c-\frac{1}{4}}$ is a real number. The
same findings than in the previous part can be made: after the passage through the critical layer, the wave is attenuated by a factor $e^{-\alpha_{l,m}\pi}$. In the {unstable} case where $\frac{l(l+1)}{m^2}\mathrm{Ri}_c < \frac{1}{4}$, $\alpha_{l,m}=\sqrt{\frac{l(l+1)}{m^2}\mathrm{Ri}_c-\frac{1}{4}}$ is a complex number, so we can not interpret the solution as upward and downward propagating waves. But the expressions given in Tab. (\ref{tab:asymp1}) and (\ref{tab:asymp2}) remain comparable to those in the first part and we deduce that the calculation of $R$ and $T$ will lead to the same result: the possibility of over-reflection and over-transmission.

\subsection{Radiative and viscous dampings}
\label{subsec:RadDamp}

\subsubsection{General equations}
We volontary left aside the factor $f(\kappa, P_{\rm r})$ in Eq. (\ref{Ui}). {In order to establish its expression, \cite{ZahnTalonMatias1997} use the equation of IGWs propagation taking into account heat diffusion but with a viscosity coefficient ($\nu$) equals to zero. Here, we generalize their result for $\nu\ne0$ ({\it i.e.} $P_{\rm r}\ne0$) and we obtain}
\begin{equation}
  f\left(\kappa,P_{\rm r}\right) = \displaystyle e^{-\tau\left(\kappa,P_{\rm r}\right)/2} \hbox{,}
  \label{tau_RV}
\end{equation}
where
\begin{equation}
\label{Eq:damping}
  \tau\left(\kappa,P_{\rm r}\right) =
  \left[l(l+1)\right]^{\frac{3}{2}}\int_{r_c}^{r_{\rm
      ZC}}\left\{\kappa\left(1+P_{\rm r}\right)\frac{N
      N_T^2}{\sigma^4}\left(\frac{N^2}{N^2-\sigma^2}\right)^{\frac{1}{2}}\frac{1}{r^3}\right\}\mathrm dr \hbox{.}
\end{equation}
{We have introduced the general expression for $N^2$, the Br\"unt-V\"ais\"al\"a frequency, to be able to take into account chemical stratification. Then, we have 
\begin{equation}N^2=N_{T}^{2}+N_{\mu}^{2}\label{BV}\end{equation} 
with $N_{T}^{2}=\frac{{\overline g}\delta}{H_{P}}\left(\nabla_{\rm
    ad}-\nabla\right)$ and $N_{\mu}^{2}=\frac{{\overline
    g}\phi}{H_{P}}\nabla_{\mu}$ where $H_{P}=\vert{\rm d}r/{\rm
  d}\ln{\overline P}\vert$ is the pressure height-scale, $\nabla=\left({\partial\ln{\overline T}}/{\partial\ln{\overline P}}\right)$ the temperature gradient and $\nabla_{\mu}=\left(\partial\ln{\overline\mu}/\partial\ln{\overline P}\right)$ the mean molecular weight ($\mu$) gradient. Moreover, we have introduced the generalized equation of state (EOS) given in \cite{KW1990}:
\begin{equation}
\frac{{\rm d}\rho}{\rho}=\frac{1}{\Gamma_1}\frac{{\rm d}P}{P}-\tilde\delta\frac{{\rm d}T}{T}+\tilde\phi\frac{{\rm d}\mu}{\mu}\,,
\label{EOS}
\end{equation}
where $\tilde\delta=-\left(\partial \ln \rho/\partial \ln
  T\right)_{P,\,\mu}$ and $\tilde\phi=\left(\partial \ln \rho/\partial \ln
  \mu\right)_{P,\,T}$.} Next, $\sigma=\sigma_{w}+m\Delta\Omega$ is the
Doppler-shifted frequency of the wave relative to the fluid rotation with
an excitation frequency $\sigma_{w}$. {$r_c$ and $r_{\rm CZ}$ are
  respectively the positions of the critical layer and of the boundary
  between the studied radiative zone and the convection region where IGWs
  are initially excited.} {This damping is another {source} of attenuation independent from the presence of a critical layer. {Moreover, as shown in \S 3.4. and \S 4., we will have to consider IGWs reflected and transmitted by unstable critical layers in addition to those initially excited by convection. Then, we introduce the following notation
\begin{eqnarray}
\label{Eq:damping_gene}
  \lefteqn{\tau\left[\kappa,P_{\rm r},r_1,r_2\right]=}\nonumber\\
  &&\left[l(l+1)\right]^{\frac{3}{2}}\int_{r_2}^{r_1}\left\{\kappa\left(1+P_{\rm r}\right)\frac{N N_T^2}{\sigma^4}\left(\frac{N^2}{N^2-\sigma^2}\right)^{\frac{1}{2}}\frac{1}{r^3}\right\}\mathrm dr \hbox{,}
\end{eqnarray}
where $r_1$ and $r_2$ are respectively the studied IGW's emission point and
the studied position with $r_1>r_2$ (in the opposite case where $r_1<r_2$,
limits in the integral have to be reversed). This will enable us to
describe radiative and viscous dampings in every configuration. Note that
in stellar radiation zones $P_{\rm r}\!<\!\!<\!1$ \citep[][]{BrunZahn2006}
inducing that the damping is mostly radiative.} We {have now to compare it with critical layers' effects}. 

\subsubsection{Progrades and retrogrades waves}
For a same environnement, prograde waves have a frequency lower than
retrograde ones (e.g. Eq. \ref{Doppler}). We choose for example a couple of
IGWs with the same excitation frequency, $\sigma_0$, same number $l$ and
opposite azimuthal degree $m$. Thus, we compare a prograde wave of
frequency $\sigma_1(r)= \sigma_{0}-\vert m \vert\Delta\Omega$ and a
retrograde one of frequency $\sigma_2(r)=\sigma_{0}+\vert
m\vert\Delta\Omega$. We obtain $\sigma_1(r)-\sigma_2(r) = -2\vert
m\vert\left(\Omega(r)-\Omega_\text{CZ}\right)<0$ in the presence of
  negative gradient of $\Omega$ and $\sigma_1(r)-\sigma_2(r) >0 $ if the
  gradient is positive. Now, let us underline that  $\tau$ given in
Eq. (\ref{tau_RV}) varies globaly as $\frac{1}{\sigma^3}$. As a
consequence, assuming that a negative $\Omega$-gradient is present
  near the excitation layer (see \S.~6), the radiative damping
is stronger for prograde waves than retrograde waves. Therefore, prograde
IGWs are absorbed by the fluid much closer to their region of excitation
while the retrograde waves are damped in a deeper region. This process
  is responsible for the net transport of angular momentum by IGWs in
  stars. On the other hand, critical layers doesnt introduce such net
bias between prograde and retrgrade IGWs because their effects scale
with $m^2$.

\subsubsection{Dependency in $l$ and $m$}
The second remark concerns the variation of {radiative and viscous dampings as a function of} $l$ and $m$. {On one hand, looking at} the multiplicative factor, which is in front of the integral in Eq. (\ref{Eq:damping}), we can roughly write {that} $\tau\propto{\frac{\left[l\left(l+1\right)\right]^{3/2}}{m^4}}$. {On the other hand,} the expression of the attenuation factor due to stable critical layers is proportional to $\left(\frac{l(l+1)}{m^2}\right)^{1/2}${, which is always greater than unity since $|m| \leq l$. The comparison between radiative and viscous dampings and the one due to stable critical layers for an IGW with given $\left(l,m\right)$ will be examined in \S 6.2.3.}

\subsubsection{Location}
Lastly, radiative and viscous dampings occur throughout the {whole} propagation of {IGWs}. On the contrary, critical layers are localised. Moreover, there is a condition for a wave to {reach} a critical layer: the rotation speed of the fluid must be of the same order than the wave frequency to hope observing $\sigma=0$. {Finally}, all IGWs are concerned by radiative and viscous dampings, {which increase around critical layers since $\tau\propto\sigma^{-4}$.}\\

In Tab. (\ref{tab:summary}), we sum up the different cases studied in these work. Depending on both properties of the fluid and of the studied wave, {these latter is submitted to} different phenomena.

\begin{table*}
\centering
\caption{Summary of the different studied cases. The observed effects depend on the properties of the wave (the frequency ($\sigma$) and degrees ($l$, $m$)) and of the fluid (thermal conductivity ($\kappa$), viscosity coefficient ($\nu$), Richardson number ($\mathrm{Ri}$)).}
\begin{tabular}{|c|c|c|c|c|c|}\cline{1-4}
\multicolumn{4}{|c|}{wave and fluid properties} &\multicolumn{2}{c}{}\\ \cline{1-4}
&&&&\multicolumn{2}{c}{}\\
$\sigma(r_c)$&$\kappa$&$\nu$&$\frac{l(l+1)}{m^2}\mathrm{Ri}_c$&\multicolumn{2}{c}{}\\
&&&&\multicolumn{2}{c}{}\\\cline{1-5}
$\ne 0 $&\multicolumn{4}{c|}{not a critical layer}&\multicolumn{1}{c}{}\\\hline
&&&&&\\
\multirow{10}{0.5cm}{\hspace{0.2cm}0}&\multirow{4}{0.5cm}{\hspace{0.2cm}0}&\multirow{4}{0.5cm}{\hspace{0.2cm}0}&$\geq
1/4$&attenuation &Fig.~\ref{fig:1}\\
&&&&&\\\cline{4-6}
&&&&&\\
&&&$< 1/4$& possible over-reflection + possible over-transmission & Fig.~\ref{fig:4} \\
&&&&&\\\cline{2-6}
&\multirow{6}{0.5cm}{\hspace{0.2cm}$\ne 0$}&&&&\multicolumn{1}{c}{}\\
&&$\ll 1$&$\geq 1/4$&attenuation + radiative damping&\multicolumn{1}{c}{}\\
&&&&&\multicolumn{1}{c}{}\\\cline{3-5}
&&&&&\multicolumn{1}{c}{}\\
&&$\ll 1$&$< 1/4$&possible over-reflection + possible over-transmission + radiative damping&\multicolumn{1}{c}{}\\
&&&&&\multicolumn{1}{c}{}\\\cline{3-5}
&&$\gg 1$&\multicolumn{2}{c|}{non stellar case}&\multicolumn{1}{c}{}\\ \cline{1-5}
\end{tabular}
 \label{tab:summary}
\end{table*}

\section{Transport of angular momentum}

{As emphasized in the introduction, our goal is to study the transport of angular momentum in stellar radiation zones and to unravel the role of critical layers. Therefore, the first step is to recall the flux of angular momentum transported by propagative IGWs and by the shear-induced turbulence. Moreover, to illustrate our purpose, we will here focus on the case of a low-mass star where IGWs are initially excited at the border of the upper convective envelope (see Fig. \ref{FigST}) by turbulent convection \citep[][]{GLS1991,Dintransetal2005,RG2005,Belkacemetal2009,Brunetal2011,LecoanetQuataert2013} and by tides if there is a close companion \citep[][]{Zahn1975,OgilvieLin2007} with an amplitude $A$ (corresponding results for massive stars with an internal convective core can be easily deduced by reversing signs).}

\begin{figure}
\begin{center}
   \includegraphics[width=0.5\textwidth]{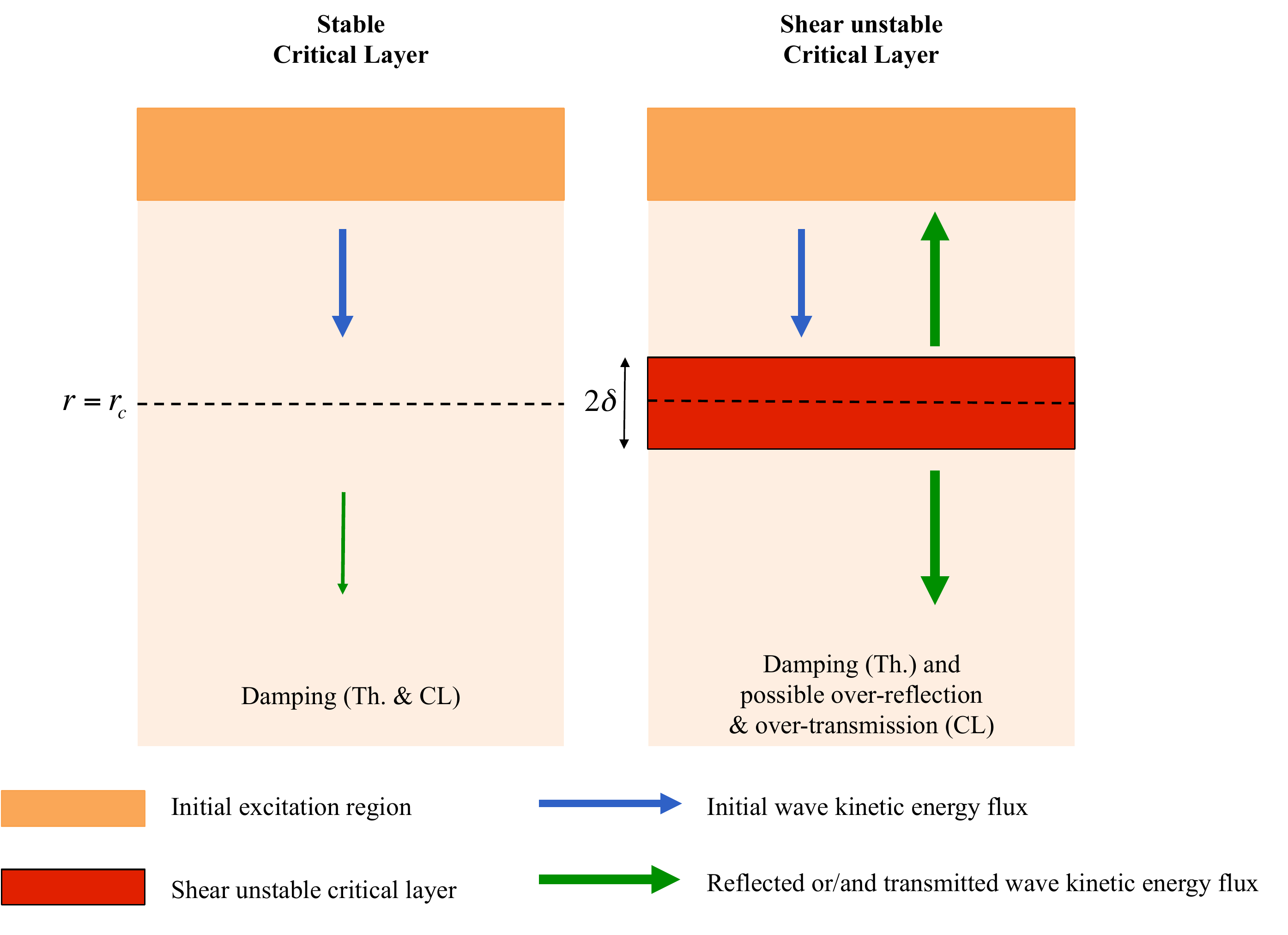}
   \caption{The two studied configurations in a low-mass star with an
     external convective envelope. {Left:} the case of a stable critical
     layer (CL) where IGWs are damped. {Right:} the case of an unstable critical layer where IGWs can be over-reflected/transmitted.} 
    \label{FigST}
\end{center} 
\end{figure} 

\subsection{Angular momentum fluxes}

\subsubsection{Angular momentum flux transported by propagative IGWs}
{First,} we have to calculate the angular momentum flux transported by a {propagative} monochromatic wave over a spherical surface. It is given by the horizontal average of the Reynolds stresses associated to the wave \citep[e.g.][]{ZahnTalonMatias1997}:
\begin{equation}
\mathcal{F}_{J;l,m,\sigma}\left(r\right)=\bar\rho r \sin{\theta} < u_{r;l,m}u_{\varphi;l,m} >_{\theta,\varphi}\hbox{,}
\end{equation}
where $< ... >_{\theta,\varphi}=\displaystyle\frac{1}{4\pi}\int_{\Omega}... \sin\theta \mathrm d\theta \mathrm d\varphi$. Besides, \cite{Mathis2009} shows that 
\begin{equation}
\mathcal{F}_{J;l,m,\sigma}= \frac{-m}{\sigma} \mathcal{F}_{E;l,m,\sigma}\left(r\right) \hbox{,}
\end{equation}
where $\mathcal{F}_{E;l,m,\sigma}$ is {the horizontal average} of the energy flux {in the vertical direction} expressed by
\cite{LighthillBook} as
\begin{equation}
\mathcal{F}_{E;l,m,\sigma}=<\hat{p}'_{l,m}u_{r;l,m} >_{\theta,\varphi} \hbox{.}
\end{equation}
So finally, the angular momentum flux is {given by}
\begin{equation}
\label{Expr:Fj}
\mathcal{F}_{J;l,m,\sigma}= \frac{-m}{\sigma} <\hat{p}'_{l,m}u_{r;l,m} >_{\theta,\varphi} \hbox{.}
\end{equation}
To calculate this angular momentum flux, we need the expressions of $u_{r;l,m}$ and $\hat{p}'_{l,m}$. $u_{r;l,m}$ is immediately accessible from Eq. (\ref{Eq:solTGS}), solution of Eq. (\ref{Eq:TGS}). It is a little more complicated for $\hat{p}'_{l,m}$ because we need to go back in the calculation leading to Eq. (\ref{Eq:TGS}). The expression of $\hat{p}'_{l,m}$ results from the first equation of the system given in Eq. (\ref{Sysp'}), which can be reduced to 
\begin{equation}
\frac{\mathrm d}{\mathrm dr}\left(\frac{\hat{p}'_{l,m}}{\bar{\rho}}\right)=\left(\sigma^2-N^2\right)\hat\xi_{r;l,m}\hbox{,} 
\end{equation}
applying the anelastic approximation and neglecting terms of order $1/L^2$.\\

{The total vertical flux of angular momentum transported by propagative IGWs is then given by
\begin{equation}
{\mathcal F}_{J}\left(r\right)=\sum\limits_{l,m,\sigma}\mathcal{F}_{J;l,m,\sigma}
\end{equation}
and we define the associated so-called action of angular momentum
\begin{equation}
{\cal{L}}_J(r)=4\pi r^2 {\mathcal F}_{J}\hbox{.}
\label{Action}
\end{equation}}

\subsubsection{Angular momentum flux transported by shear-induced turbulence}

{In the case of shear-unstable regions, such as region II in the regime where $\mathrm{Ri}_c \geq \frac{1}{4}\frac{m^2}{l\left(l+1\right)}$, IGWs are unstable \citep[e.g.][]{DrazinReid2004}. After the non-linear saturation of the instability, a steady turbulent state is reached. Then, as established for example in \cite{Zahn1992,TalonZahn1997}, and now confirmed by numerical simulations \citep[see][]{PratLignieres2013}, the vertical flux of angular momentum transported by shear-induced turbulence is given by
\begin{equation}
{\mathcal F}_{T;V}\left(r\right)={\overline\rho r^2 \nu_{V}\left(r\right)\partial_r{\overline\Omega}},
\label{turbuVF}
\end{equation}
where
\begin{equation}
\nu_{V}= {{\rm Ri}^{\rm c} \over N^2_{T} /(\kappa + \nu_{H}) + N^{2}_{\mu}/\nu_{H}}
\left(r\partial_{r}\overline\Omega\right)^2\,. 
\label{eq:dv-mu}
\end{equation}
${\rm Ri}^{\rm c} = 1/6$ is the adopted value for the critical Richardson number and $\nu_H$ is the horizontal turbulent viscosity for which we assume the prescription derived by \cite{Zahn1992}.}

\subsubsection{Equation of transport of angular momentum}
Let us now refocus {these} results in the wider frame of {the complete} angular momentum transport theory. Considering the other transport {mechanisms} we presented in introduction, the angular momentum transport equation {taking into account meridional flows, shear-induced turbulence and IGWs \citep[e.g.][]{TalonCharbonnel2005,Mathis2009} becomes}
\begin{eqnarray}
  {\overline\rho}\frac{\mathrm d}{\mathrm dt}\left(r^2{\overline\Omega}\right)&=& 
   \frac{1}{5r^2}\partial_r\left({\overline\rho} r^4 {\overline\Omega} U_2\left(r\right)\right)\nonumber\\ 
  &&+\begin{cases}
  -\displaystyle\frac{1}{4\pi r^2}\partial_r{\cal{L}}_J\left(r\right)\hbox{where IGWs are propagative}\\
  \hbox{}\\
  \displaystyle\frac{1}{r^2}\partial_r\left(r^2{\mathcal F}_{T;V}\left(r\right)\right)\hbox{for unstable regions.}\\
  \end{cases}
\end{eqnarray}
{The first term in the right hand side corresponds to the advection of angular momentum by the meridional circulation, where $U_r = U_2\,P_2(\cos\theta)$ is its radial component. $\frac{\mathrm d}{\mathrm dt}=\frac{\partial}{\partial t}+\dot{r}\frac{\partial}{\partial r}$ is the Lagrangian derivative that takes into account the radial contractions and dilatations of the star during its evolution, which are described by $\dot{r}\vec{\hat e}_r$. According to our hypothesis, we do not take into account the transport by the Lorentz force, associated with magnetic fields in stellar radiative zones \citep{MathisZahn2005}.}\\

{The major difference with previously published equations is that IGWs
  and shear-induced turbulence transports of angular momentum are not
  summed linearly since they are, as we demonstrated before, intrisically
  coupled. Therefore, for stable regions, one must take into account IGWs'
  Reynolds stresses (Eq. \ref{Action}) only, while for unstable regions, only the vertical turbulent flux given in Eq. (\ref{turbuVF}) must be taken into account. Let us now examine each case of critical layer.}

\subsection{Stable critical layer $\left(\mathrm{Ri}_c \geq \displaystyle\frac{1}{4}\frac{m^2}{l\left(l+1\right)}\right)$}

\subsubsection{Case of the perfect fluid}
{This is the simplest case of a stable critical layer in a perfect fluid. Then, we apply Eq. (\ref{Expr:Fj}) to obtain the expressions for the transported fluxes by propagative IGWs below and above the critical layer:
\begin{equation}
\label{eq:FP}
  \left\{
      \begin{array}{l}
\mathcal{F}_{J;l,m,\sigma}\left(r\ge r_c\right)=\displaystyle \frac{1}{r^2}\frac{1}{2}m A^2\,\frac{{\mathcal J}_{l,m}}{l\left(l+1\right)}\\
\\
\mathcal{F}_{J;l,m,\sigma}\left(r<r_c\right)=\displaystyle \frac{1}{r^2}\frac{1}{2}m A^2\,\frac{{\mathcal J}_{l,m}}{l\left(l+1\right)}e^{\displaystyle-2\pi\alpha_{l,m}}\hbox{,}
      \end{array}
  \right.
\end{equation}
where A is the initial amplitude of the IGW at $r=r_{\rm CZ}$, $\alpha_{l,m}=\sqrt{\frac{l(l+1)}{m^2}\mathrm{Ri}_c-\frac{1}{4}}$ (see Eq. (\ref{eq:alphalm})) and
\begin{eqnarray}
\lefteqn{{\mathcal J}_{l,m}=<\left[{P}_{l}^{m}\left(\cos\theta\right)\right]^2>_{\theta}=\frac{1}{2}\int_{0}^{\pi}\left[{P}_{l}^{m}\left(\cos\theta\right)\right]^2\sin\theta{\rm d}\theta}\nonumber\\
&=&\frac{2}{2l+1}\frac{\left(l+\vert m \vert\right)!}{\left(l - \vert m\vert \right)!}\hbox{,}
\end{eqnarray} 
where ${P}_{l}^{m}$ are the associated Legendre polynomials. \cite{BookerBretherton1967} have obtained similar results in cartesian coordinates. Let us make two remarks about these expressions. First, we see the expected attenuation of the flux by a factor $e^{-2\pi\alpha_{l,m}}$ when the wave passes through the critical layer. Moreover, $\mathcal{F}_{J;l,m,\sigma\pm}$ depends on $m$ (and not on $m^2$). As a consequence, we recover the classical result that prograde waves ($m<0$) and retrograde ones ($m>0$) have opposite angular momentum flux (respectively a deposit and an extraction). Finally, the monochromatic action of angular momentum ${\cal{L}}_{J;l,m,\sigma}(r)=4\pi r^2 {\mathcal F}_{J;l,m,\sigma}$ is constant in each region because of the absence of dissipation.}

\subsubsection{Case of the non-perfect fuid}
We saw in the previous part that the solution of the equation of propagation in the case of a non-perfect fluid is similar to the one obtained for a perfect fluid. For this reason, we are allowed to apply the same method for the calculation of $\mathcal{F}_{J;l,m,\sigma} $ and we obtain
\begin{equation}
\label{eq:FNP}
  \left\{
      \begin{array}{l}
\mathcal{F}_{J;l,m,\sigma}\left(r\ge r_c\right)=\\
\displaystyle \frac{1}{r^2}\frac{1}{2}m A^2\,\frac{{\mathcal J}_{l,m}}{l\left(l+1\right)}e^{\displaystyle-\tau\left[\kappa,P_{\rm r}, r_{\rm CZ},r\right]}\\
\\
\mathcal{F}_{J;l,m,\sigma}\left(r<r_c\right)=\\
\displaystyle \frac{1}{r^2}\frac{1}{2}m A^2\,\frac{{\mathcal J}_{l,m}}{l\left(l+1\right)}e^{\displaystyle-2\pi\alpha_{l,m}}e^{\displaystyle-\tau\left[\kappa,P_{\rm r}, r_{\rm CZ},r\right]}\hbox{.}
      \end{array}
  \right.
\end{equation}
The difference with Eq. \eqref{eq:FP} comes from the introduction of radiative {and viscous} dampings (Eq. \ref{Eq:damping_gene}).\\ 

The conclusion is that, in the stable case, the attenuation due to the passage through a critical layer is added to {dampings due to dissipation.} This observation will lead us to simply implement the role of stable critical layers as an additional term in the damping coefficient.

\subsection{Unstable critical layer $\left(\mathrm{Ri}_c \leq \displaystyle\frac{1}{4}\frac{m^2}{l\left(l+1\right)}\right)$}

\subsubsection{Region I: $r_c+\delta \le r \le r_{\rm CZ}$}
{Using the summary given in Fig. \ref{FigST} for the unstable case and the obtained results for regions where IGWs are propagative (\S 5.2.2.), we obtain:}
\begin{eqnarray}
\lefteqn{\mathcal{F}_{J;l,m,\sigma}\left(r_c+\delta \le r \le r_{\rm CZ}\right)=}\nonumber\\
&&\frac{1}{r^2}\frac{1}{2}m A^2\,\frac{{\mathcal J}_{l,m}}{l\left(l+1\right)}e^{\displaystyle-\tau\left[\kappa,P_{\rm r}, r_{\rm CZ},r\right]}\nonumber\\
&+&\frac{1}{r^2}\frac{1}{2}m \left[A^2{\vert R \vert}^{2}e^{\displaystyle-\tau\left[\kappa,P_{\rm r}, r_{\rm CZ},r_c+\delta \right]}\right]\,\frac{{\mathcal J}_{l,m}}{l\left(l+1\right)}\nonumber\\
&&\times\,e^{\displaystyle-\tau\left[\kappa,P_{\rm r}, r ,r_c+\delta \right]}\hbox{,}
\end{eqnarray}
{where we identify the transport induced by the incident wave, which propagates downward, and the one induce by the reflected one, which propagates upward. Then, we can see that the unstable critical layer is a second excitation source for IGWs propagating in region I and that the angular momentum transport will be modified in such situation, particularly when over-reflection ($\vert R \vert>1$) occurs.}

\subsubsection{Region II: $r_c-\delta<r<r_c+\delta$}

{In this unstable region, we directly use results obtained in \S 5.1.2. to describe the flux of angular momentum transported by the shear-induced turbulence, i.e.:}
\begin{equation}
{\mathcal F}_{T;V}\left(r_c-\delta<r<r_c+\delta\right)={\overline\rho r^2 \nu_{V}\partial_r{\overline\Omega}}.
\label{turbuVF}
\end{equation}

\subsubsection{Region III: $r\ge r_c-\delta$}
{Using the same methodology that for region I, we get in a straightfoward way:}
\begin{eqnarray}
\lefteqn{\mathcal{F}_{J;l,m,\sigma}\left(r\ge r_c-\delta\right)=}\nonumber\\
&&\frac{1}{r^2}\frac{1}{2}m \left[A^2{\vert T \vert}^{2}e^{\displaystyle-\tau\left[\kappa,P_{\rm r}, r_{\rm CZ},r_c+\delta \right]}\right]\,\frac{{\mathcal J}_{l,m}}{l\left(l+1\right)}\nonumber\\
&&\times\,e^{\displaystyle-\tau\left[\kappa,P_{\rm r}, r_c-\delta, r \right]}\hbox{,}
\end{eqnarray}
{where we indentify the transport induced by the transmitted wave, which propagate downward. As in region I, we can see that the unstable critical layer constitutes a secondary excitation source for IGWs propagating in region III and that the angular momentum transport will be modified, particularly when over-transmission ($\vert T\vert>1$) occurs.}\\

{Since the general theoretical framework has been given, we have now to explore the possibility of the existence of the two different regimes (stable and unstable) along the evolution of a given star. As a first application, we choose to study the case of a solar-type star which has already been studied without critical layers by \cite{TalonCharbonnel2005}.}

\section{A first application: the evolution of a solar-type star}

\subsection{The STAREVOL code}

We use the one dimensional hydrodynamical Lagrangian stellar evolution code STAREVOL (V3.10), and the reader is referred to \citet[]{Lagardeetal2012} and references therein for a detailed description of the input physics. We simply recall the main characteristics and parameters used for the modelling that are directly relevant for the present {work}. We use the Scharwschild criterion to determine convective zones position, and their temperature gradient is computed according to the MLT with a $\alpha_\text{MLT} = 1.75$. The solar composition is taken from \citet{AsplundGrevesse2005} with the Ne abundance from
\citep{CunhaHubeny2006}. We have generated the opacity table for temperature higher than 8000~K following \citet{IglesiasRogers1996} by using their website\footnote{\url{http://adg.llnl.gov/Research/OPAL/opal.html}}. Opacity table at lower temperature follows \citet{FergusonAlexander2005}\footnote{\url{http://webs.wichita.edu/physics/opacity/}}. The mass loss rate is determined following \citet{Reimers1975} with a parameter $\eta_R = 0.5$. The increase of mass loss due to rotation is taken into account following \citet{MaederMeynet2001}. However due to the small mass loss and velocity of our model this effect remains weak.\\

In radiative regions, we follow \citet{MathisZahn2004} formalism for the
transport of angular momentum and of chemicals as well as the prescription
from \citet{TalonZahn1997} for the vertical turbulent transport
(Eq. \ref{eq:dv-mu}). We assume that convective regions rotate as
solid-body. The treatment of IGW follows
\citet{TalonCharbonnel2005,TalonCharbonnel2008} with the difference that
the volumetric excitation by Reynolds stresses in the bulk of convective
  zones \citep[e.g.][]{1990ApJ...363..694G,1994ApJ...424..466G,Belkacemetal2009} is consistently computed at each time-step as a function of their
physical properties.\\

We start with an initial model of 1.0~\Ms{} at solar metallicity, with an
initial surface velocity of 70~\kms. The rotation profile is initially
flat. We add magnetic braking through the following
law: $\frac{\mathrm{d}J}{\mathrm{d}t} = - K \Omega^4$ with a constant $K = 3\times 10^{30}$. This value has been calibrated to reproduce the
surface velocity determined in the Hyades by \citep{Gaige1993}.

\subsection{The effects of critical layers}

\subsubsection{Location of critical layers}

We theoretically studied the impact of the critical layer for a given IGW,
thus assuming that there are some radii where the relation $\sigma(r)=0$ is
satisfied. As a consequence, the first question we may answer thanks to the
simulation concerns the existence of such critical layers and their
location in the radiative zone. Figure \ref{fig:6} shows that critical
layers do exist in the studied solar-like star's radiative core. In the three panels, we superimpose the rotational velocity of the star's interior as a function of the normalized radius with the position of potential critical layers, marked out with colorized squares {which correspond to positions where $\sigma=\sigma_{w}+m\Delta\Omega=0$.} Each panel corresponds to a {given} value of the excitation frequency $\sigma_{w}$. As expected, the positions of the critical layers only depend on the azimuthal number $m$ of the wave, and not on the degree $l$. Thanks to this plot, we {confirm} that critical layers exist in the {studied} solar-like star. However, we have not already taken into account the fact that all waves {cannot reach these positions}.
\begin{figure}
  \centering
  \includegraphics[width=0.5\textwidth]{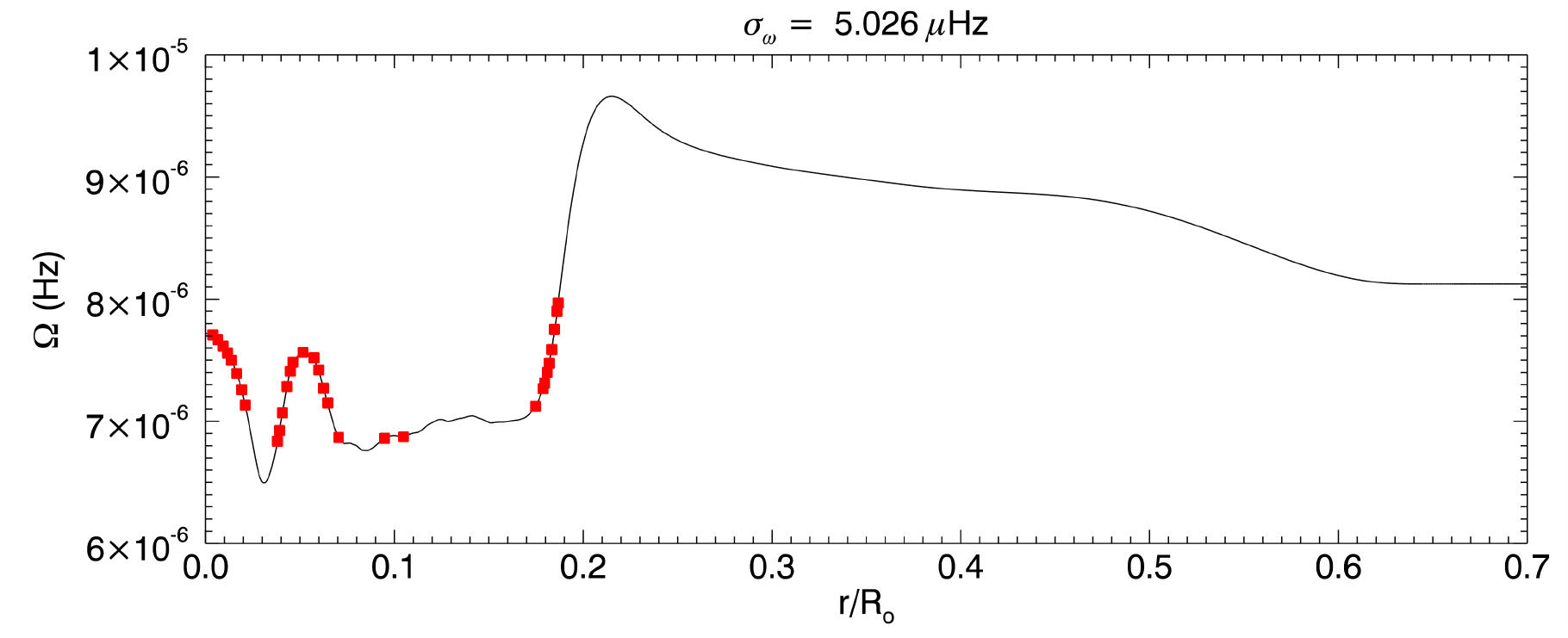}
\includegraphics[width=0.5\textwidth]{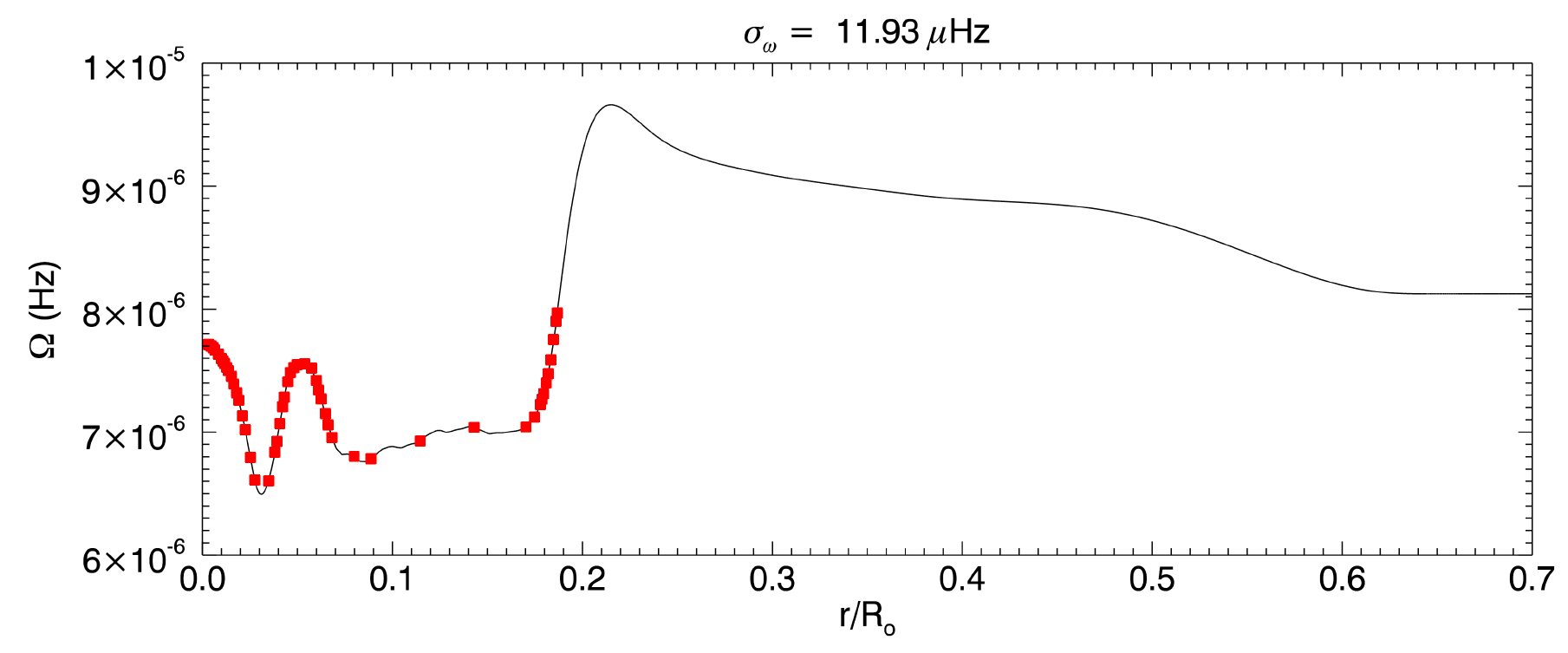}
\includegraphics[width=0.5\textwidth]{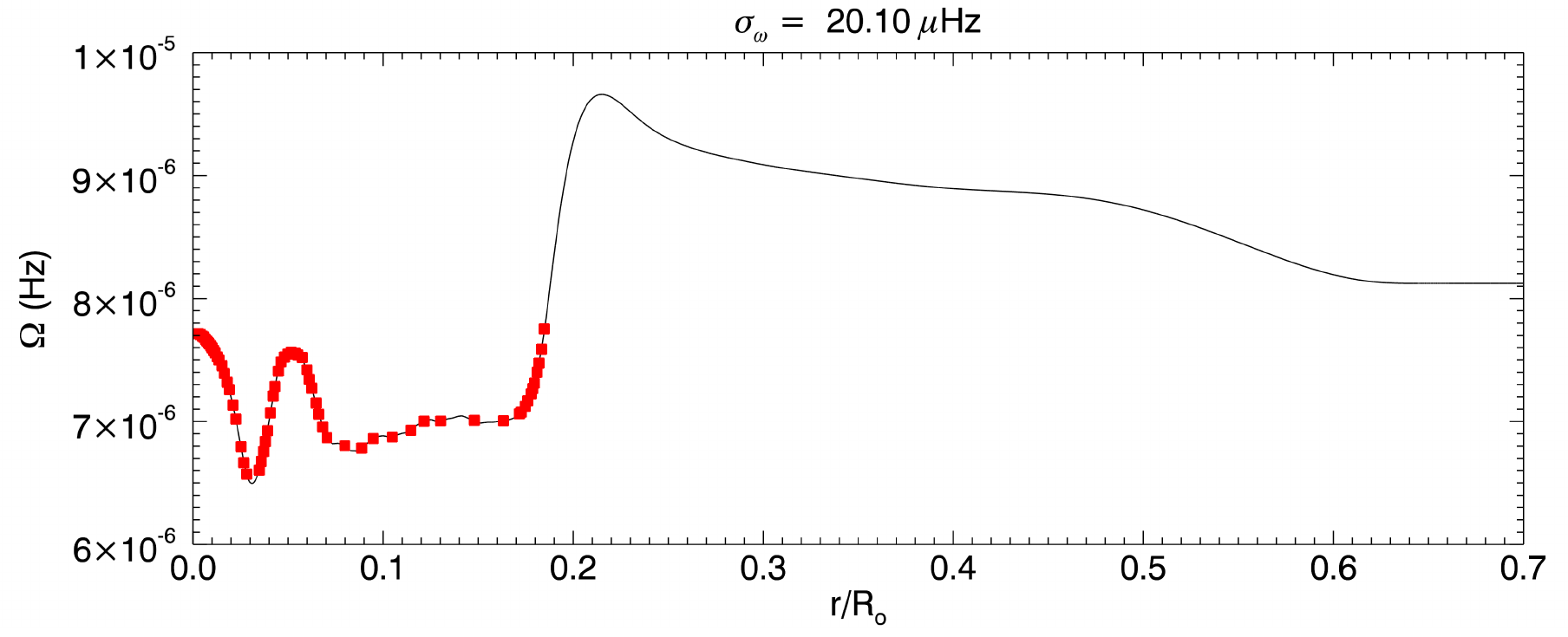}
  \caption{Position of the critical layers for retrograde IGWs for three different frequencies, superimposed with the rotational velocity profile in the star.}
  \label{fig:6}
\end{figure}\\

First of all, we only {take into account} retrograde waves {because the}
prograde {ones are} damped immediately after their excitation (see
\S~4.4.2). Moreover, in the left panel of Fig.~\ref{fig:7}, we have
represented in logarithmic scale the luminosity of a given IGW at the
moment of its initial stochastic excitation by the turbulent convection as
a function of its degrees $l$ and $m$, following the spectrum adopted in \cite{TalonCharbonnel2005}. The difference between the three plots is the
value of the excitation frequency $\sigma_{w}$. Thanks to this representation, we can see that the maximum of excitation lies in a domain where $l$ and $m$ are close and quite small. Moreover, it shows that for each excitation frequency, the amplitude of the excited wave depends on $l$ and $m$, and may be close to zero. As a consequence, some critical layers represented in Fig.~\ref{fig:6} may belong to a non-excited wave, or to a wave completely damped at this depth. Fortunately, obtained results show that some waves really meet their critical layer.
\begin{figure*}
  \includegraphics[width=0.45\textwidth]{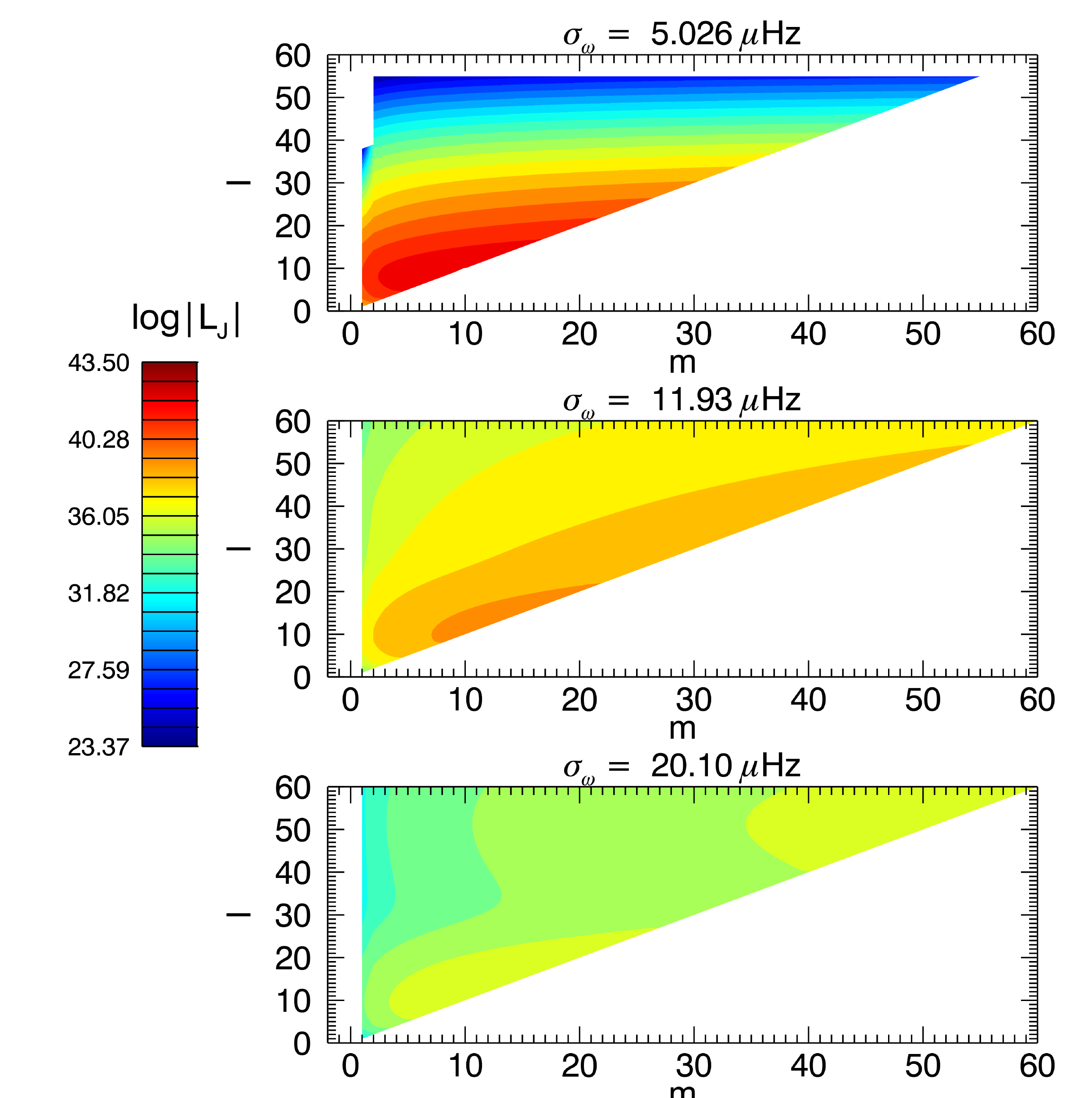}
 \includegraphics[width=0.45\textwidth]{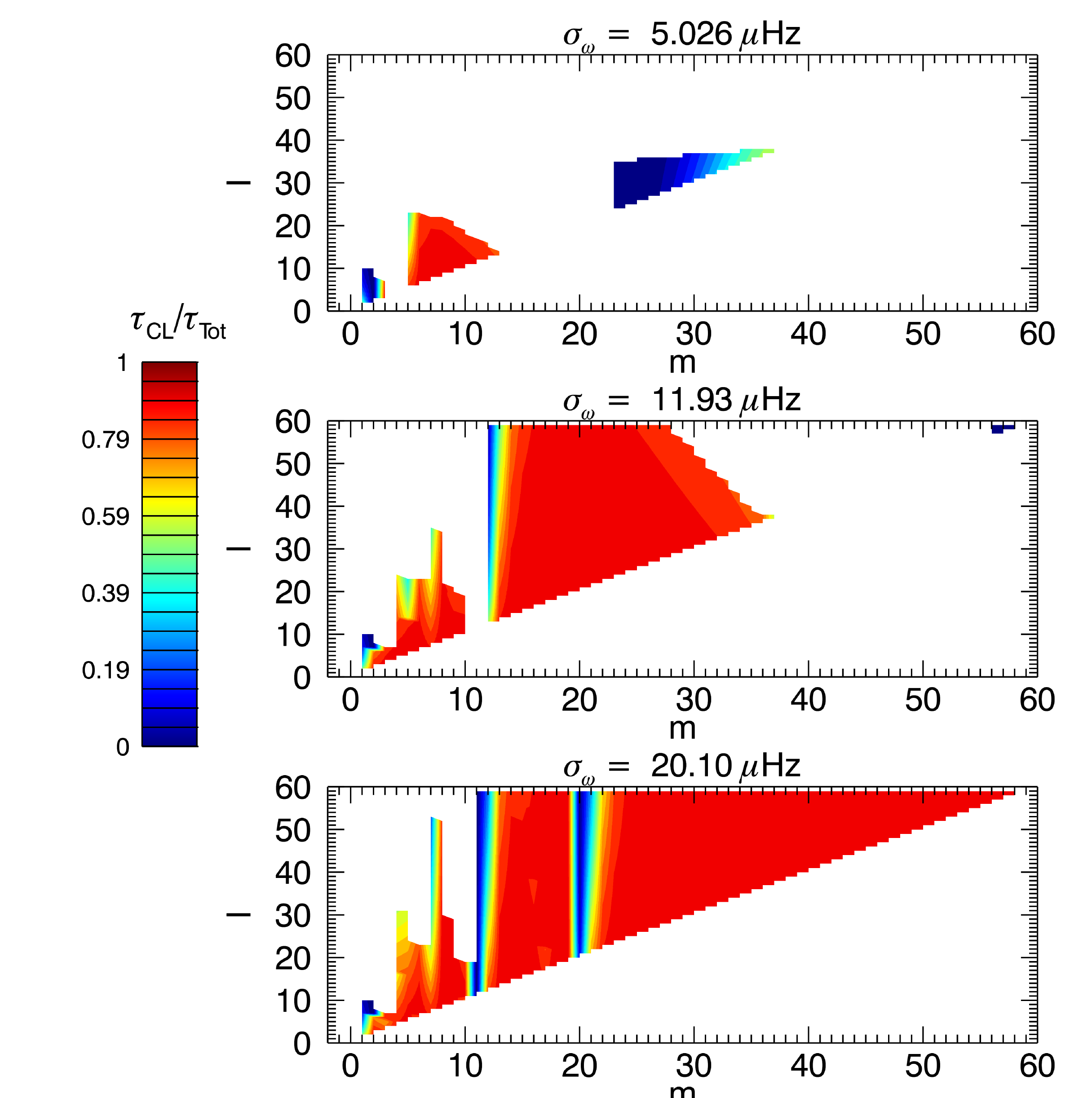}
  \caption{Left: Luminosity of the waves at the location of their
    excitation as function of $l$ and $m$. Right: Ratio between $\tau_{CL}$
    and $\tau_{tot} = \tau_{CL} +\tau_{\mathrm{rad}} $ (see
    Eq.~\eqref{eq:3}).}
  \label{fig:7}
\end{figure*}
 
\subsubsection{Interaction between waves and critical layers}
Concerning the way these latters interact with the surrounding fluid, the theory predicts two possible regimes depending on the value of the Richardson number $\mathrm{Ri}_c$ at the
critical level. In our simulation, it appears that for every detected critical layer, the relation $\mathrm{Ri}_c > \frac{1}{4}\frac{m^2}{l\left(l+1\right)}$, which correponds to the stable regime, is verified. As a consequence, we establish that the second unstable regime (with the associated possible tunneling or over-reflection and transmission) does not occur for the solar-like star of our simulation; forthcoming studies may explore other types of stars at different evolutionary stages, to see if this regime can occur. Therefore, for the solar-type star studied here, we only implement in STAREVOL the terms related to stable critical layers: each time a wave passes through a critical layer, it is thus damped with a coefficient $e^{-\pi\sqrt{\frac{l(l+1)}{m^2}\mathrm{Ri}_c-\frac{1}{4}}}$ (see Eq. \ref{eq:FNP}).

\subsubsection{Effect of critical layers}
Since all interactions between waves and critical layers are of the same kind in the studied star, we can concentrate on the quantitative importance of their effect on the transport of angular momentum. We know that in the stable regime (\S 5.2.2.), the wave passing through its critical layer is damped by the factor given in Eq. \ref{eq:FNP}) that is added to radiative damping (here $P_{\rm r}\!<\!\!<\!1$ and the viscous damping is thus negligible), which has already been taken into account in previous works \citep[e.g.][]{TalonCharbonnel2005}. In the right panel of Figure \ref{fig:7}, we thus choose to represent the ratio between $\tau_{\rm CL}$, the rate of attenuation due to the passage of the wave through its stable critical layer and the sum $\tau_{\rm tot}$ of this rate and the one of the radiative damping. The explicite formula is
\begin{eqnarray}
  \label{eq:3}
  \lefteqn{\frac{\tau_{\rm CL}}{\tau_{\rm tot}}=}\\
  &&\frac{2\pi\sqrt{\frac{l(l+1)}{m^2}\mathrm{Ri}_c-\frac{1}{4}}}{2\pi\sqrt{\frac{l(l+1)}{m^2}\mathrm{Ri}_c-\frac{1}{4}}\!+\!\left[l\left(l+1\right)\right]^{\frac{3}{2}}\displaystyle\int_{r_c}^{r_\mathrm{zc}}{\kappa \frac{N N_T^2}{\sigma^4}\left(\frac{N^2}{N^2-\sigma^2}\right)^{\frac{1}{2}}\frac{1}{r^3}\mathrm{d}r}}\hbox{.}\nonumber
\end{eqnarray}

On the contrary of the left panel, here are represented only waves which meet a critical layer. That is the reason why several white zones are seen. They correspond to the waves which have been attenuated before reaching the depth of their critical layer. In the case of low $\sigma_w$ (upper panels in Fig. \ref{fig:7}), the high degree waves ($l>45$) are simply not excited, as we can see on the right. In red zones, the role of critical layers is important in
comparison with the radiative damping while dark blue regions are those where the radiative damping dominates . Therefore, this figure shows that critical layers should be taken into account.

\subsubsection{Evolution of the rotation profile}

\begin{figure}
  \centering \includegraphics[width=0.45\textwidth]{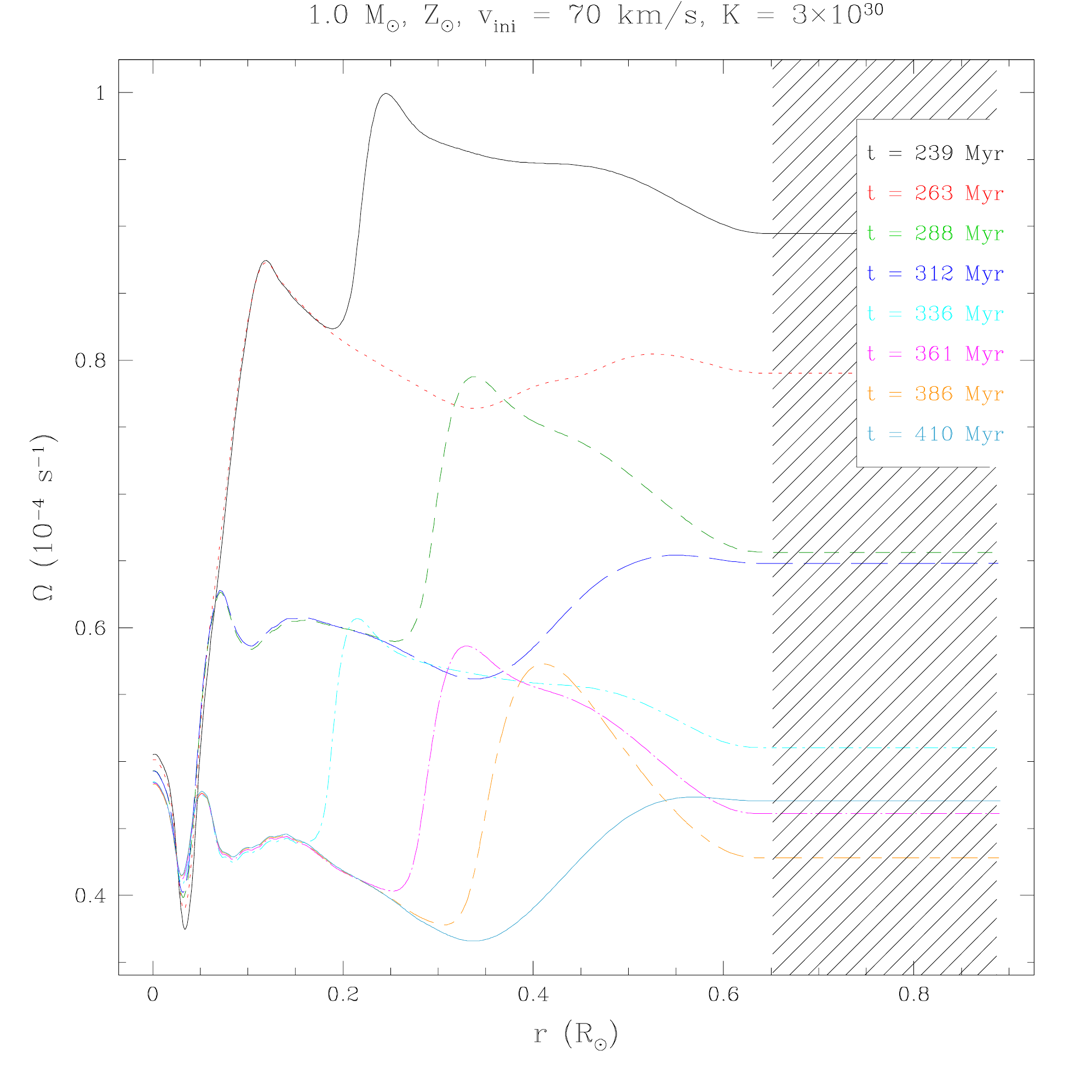}
  \includegraphics[width=0.45\textwidth]{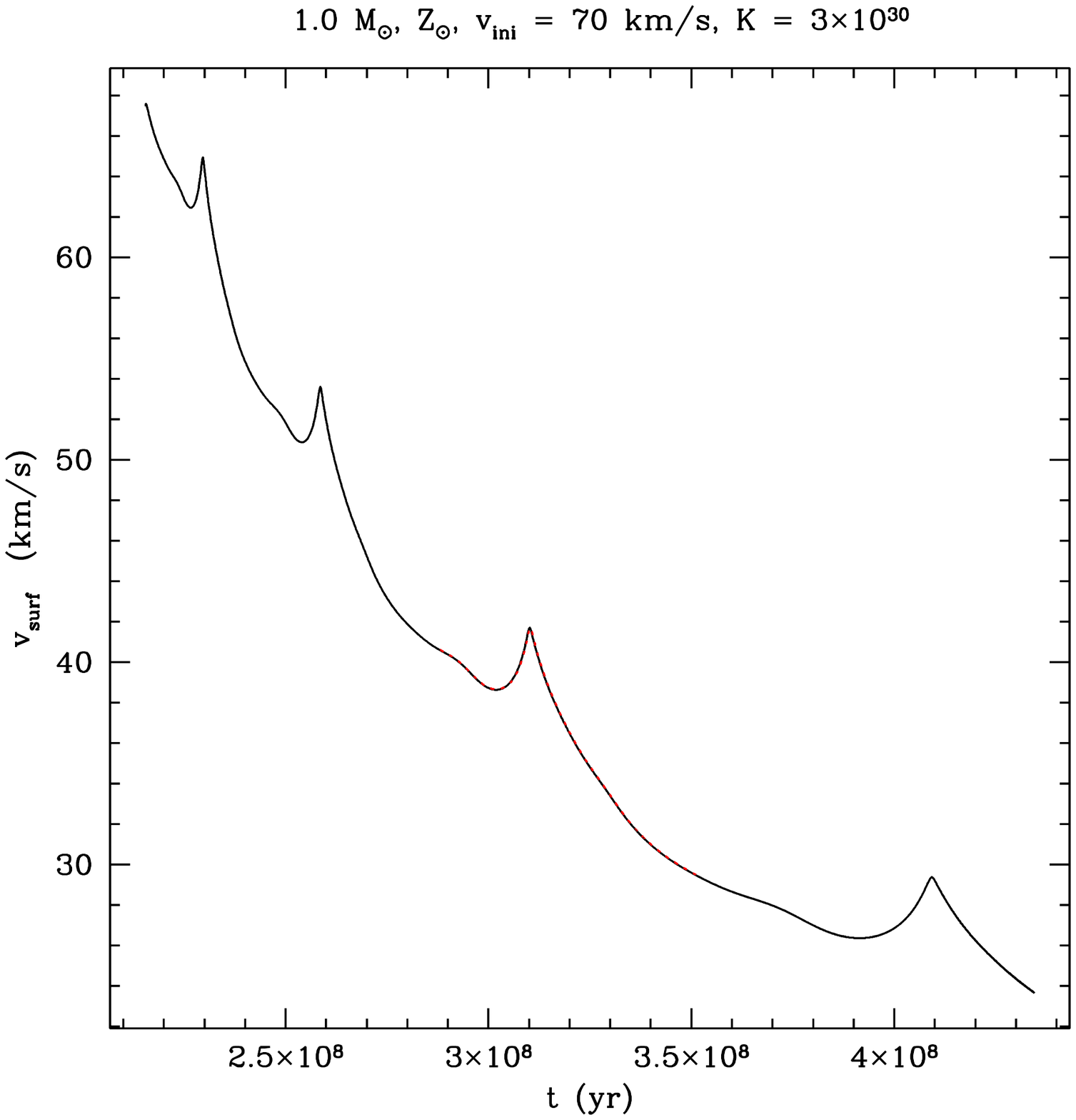}
  \caption{Evolution of the rotation profile where the role of the critical
    layers in the transport of angular momentum is taken into
    account. The curves are labelled according to the corresponding ages in
    Gyr. Parameters are indicated above where K is the 
    braking constant. Right : Comparison between the evolution of the rotation with (red)  and
    without (black) taking into account the effect of the critical layers.}
  \label{fig:8}
\end{figure}

Let us now concentrate on the evolution of the rotation profile when momentum deposition due to the interaction between waves and critical layers is taken into account. The other transport mechanism (IGWs' radiative damping effects, the shear-induced turbulence and the meridional circulation) have been previously implemented in the code \citep{TalonCharbonnel2005}.\\

In the center of the star, the rotation velocity is lower and increases the
radiative damping of retrograde waves (see Eq. \ref{Eq:damping}). This
forms an angular momentum extraction front which propagates from the core
to the surface to damp the differential rotation. Three fronts are seen in
the top panel of Fig.~\ref{fig:8} and have the same form that those already
obtained in \cite{TalonCharbonnel2005}. To isolate the action of critical
layers on the evolution of the rotation profile, we have superimposed in
the bottom panel of Fig. \ref{fig:8} the surface velocity as function of
the evolution time with (black line) and without (red line) critical layer
effects between $2.8 \times 10^8$ and $3.5\times 10^8$ years. Both curves
are nearly identical wich shows that despite their local action showed in
Fig.~\ref{fig:6}, IGWs' stable critical layers do not disturb the dynamical
evolution of surface velocity in the case of the studied star. This can be
easily understood since the radiative damping becomes mostly efficient
around critical layers' positions because of its dependance on
$\sigma^{-4}$. Moreover, if in \cite{TalonCharbonnel2005}, the IGWs action
and the shear-induced turbulence have been added as uncoupled physical
mechanisms, it has been demonstrated that in the upper region ($r>r_c$)
where IGWs are propagative, the coefficient $\nu_V$ is negligible, while in
the inner one ($r<r_c$) the differential rotation has been damped, leading
to the same result with a transport dominated by the meridional circulation.\\

This clearly indicates that unstable critical layers will lead to major
modification of the transport of angular momentum in stellar interiors. A
systematic exploration of different type of stars for different
evolutionary stages will be undertaken in a near future to explore their
possible existence.
Moreover, in order to give quantitative informations, it is necessary to improve
the way waves are excited in this model. A future work will implement a
prescription about penetrative convection processes. The best way to do it
is to use a realistic numerical simulation of such mechanism to obtain the
excitation spectrum at the base of the convective zone \citep{RG2005,
  Brunetal2011, 2012sf2a.conf..289A}.

\section{Conclusion}

In this paper, we {study in details} a new mechanism of interaction between IGWs and the shear {of the mean flow that occurs at co-rotation layers in stably stratified stellar radiation zones.} Taking advantage of the work realized in the {literature} about atmospheric {and oceanic} fluids, we highlight the similarities with {such stellar regions} and propose an analytical
approach adapted to {the related case of deep spherical shells}. Then, the use of spherical coordinates brings differences in the equations and make their resolution more complicated but the final results are comparable. We {then demonstrate the intrinsic couplings between IGWs and the shear-induced instabilities and turbulence that can thus not be added linearly as done previously in stellar evolution literature. Then,} we highlight the existence of two regimes where the interactions between IGWs and the {shear at critical layers are strongly different: 
\begin{itemize}
\item in the first case, the fluid is stable and IGWs amplitude is overdamped by the critical layer compared to the classical case where only radiative and viscous dampings are taken into account;
\item in the second case, the fluid is unstable and turbulent and the
  critical layer acts as a secondary excitation region. Indeed, through
  over-reflection/transmission (when $\vert R\vert>1$ and $\vert T\vert>1$)
  energy is taken from the unstable shear that increase the amplitude of an
  incident IGW. Moreover, even in the case of simple reflection and
  transmission where $\vert R\vert<1$ and $\vert T\vert<1$, this
  demonstrate the existence of IGWs "tunneling" through unstable regions as
  identified by \cite{SutherlanYewchuk2004} in laboratory experiments and by \cite{BrownSutherland2006,NaultSutherland2007} in Geophysics.
\end{itemize}} 
Therefore, these mechanisms opens a {new} field of investigations concerning angular momentum tranport processes by IGWs in stellar interiors.\\ 

{Indeed, even if} according to {our first} evolutionary calculations {with STAREVOL}, only the first stable regime exists in solar-type stars, we expect to find stars {where the unstable regime and possible tunneling or over-reflection/transmission take place. Moreover,} while the formalism presented in this work is general, several uncertainties remain. First, concerning the {stochastic excitation} of IGWs by convection, the model used in our evolutionnary code certainly underestimates the wave flux since it considers only {the volumetric} excitation {in the bulk of the convective envelope} while convective penetration {should be also taken into account}. This will influence the measured action of
critical layers since it is proportional to the {initial IGWs} amplitude. Then, only retrograde waves are simulated here, considering that prograde ones are immediately damped and do not penetrate deeply in the radiation zone. This should normaly not affect our results because the critical layers we detected are located in the {deep} radiation zone, but formal equations take both types of waves into account.\\

The last point to bear in mind is that no latitudinal dependence for the
angular velocity is considered here. We explained the reason of this choice
in the introduction but one must not forget this approximation. {Finally,
  since our goal is to get a complete and coherent picture of the transport
  of angular momentum in stellar radiation zones for every stellar type or
  evolutionary stage, it will be important to extend this work to the cases of gravito-inertial waves, where the action of
  Coriolis and centrifugal accelerations is considered
  \citep[e.g.][]{LeeSaio1997,DintransRieutord2000,Mathis2009,Ballotetal2011}
  and of magneto-gravito-inertial waves, when radiation zones are magnetized
  \citep[e.g.][]{RV1972,KMG2003,MGR2011,MdB2012}.}

\section*{Acknowledgments}
The authors thanks the referee for her/his comments that allowed to
 improve the paper. This work was supported by the French Programme
National de Physique Stellaire (PNPS) of CNRS/INSU, by the CNES-SOHO/GOLF
grant and asteroseismology support in CEA-Saclay, by the Campus Spatial of
the University Paris-Diderot and by the TOUPIES project funded by the French National Agency for Research (ANR). S. M. and L. A. are grateful to Geneva Observatory where part of this work has been achieved. T.D. acknowledges financial support from the Swiss National Science Foundation (FNS) and from ESF-Eurogenesis.

\section*{Appendix A: Validity of the JWKB approximation}
The form of the equation to solve is: 
\begin{equation}
\label{forme_TGS}
\frac{\mathrm d^2 \Psi}{\mathrm dr^2} = f\left(r\right)\Psi\left(r\right)\hbox{.}
\end{equation}
The first step is to introduce the Liouville transformation \citep[e.g.][]{Olver1974}, i.e.:
\begin{equation}
W\left(r\right) = f^{1/4} \Psi \hbox{, and } \xi\left(r\right) = \int^{r} f^{1/2} {\rm d}r'\hbox{.}
\end{equation}
We deduce : 
\begin{eqnarray}
\frac{\mathrm dW}{\mathrm d\xi} &=& \frac{1}{4}f^{-5/4}f'\Psi+f^{-1/4}\Psi' \hbox{,} \\
\frac{\mathrm d^2W}{\mathrm d\xi^2} &=&-\frac{5}{16}f^{-11/4}f'^2\Psi+\frac{1}{4}f^{-7/4}f''\Psi+f^{-3/4}\Psi'' \hbox{,}
\end{eqnarray}
and Eq. (\ref{forme_TGS}) becomes
\begin{equation}
\frac{\mathrm d^2W}{\mathrm d\xi^2}=\left[1+\Phi\left(r\right)\right]W \hbox{,}
\end{equation}
where $\Phi = \displaystyle\frac{4ff''-5f'^2}{16f^3}$.\\
In the present case we have $f\left(r\right) = -k_V^2\left(r\right) = -\displaystyle\frac{l(l+1)}{r^2}\left(\displaystyle\frac{N^2}{\sigma^2}-1\right)$. The WKBJ approximation is available when $N^2\gg\sigma^2$. Consequently, we get
\begin{eqnarray}
&& f\approx-l(l+1)\frac{1}{r^2}\frac{N^2}{\sigma^2} \hbox{,}\\
&& f'\left(r\right)\approx l(l+1)\frac{2N^2}{r^2 \sigma^3}m\bar{\Omega}' \hbox{,}\\
&& f''\left(r\right)\approx -l(l+1)6m^2\frac{N^2\bar{\Omega}'^2}{r^2 \sigma^4} \hbox{,}
\end{eqnarray}
and
\begin{equation}
\Phi(r)\approx-\frac{1}{4}\frac{m^2}{l(l+1)}\frac{\left(\frac{\mathrm{d}\bar\Omega}{\mathrm{d}r}\right)^2r^2}{N^2}\equiv-\frac{1}{4}\frac{m^2}{l(l+1)}{\mathrm R_{i}}\hbox{,}\\
\end{equation}
where $\mathrm{Ri}$ is the Richardson number of the fluid defined in Eq.\eqref{eq:Ri}.\\

At last, we obtain that the condition for applying the WKBJ approximation is $|\Phi| \ll 1$ which leads to $\mathrm{Ri} \gg \displaystyle\frac{1}{4}\frac{l(l+1)}{m^2}$.

\section*{Appendix B: Mathematical treatment for the non-perfect fluid case}
Let us introduce
\begin{eqnarray}
\chi_{l,m}&=&\int_{a}^{b}{e^{\eta t}v\left(t\right)dt}\hbox{,}
\end{eqnarray}
where $a$ and $b$ are the limits of a domain which will be defined later. The equation of propagation can then be written
\begin{equation}
\frac{1}{P_{\mathrm r}}t^6v+\left(1+\frac{1}{P_{\mathrm r}}\right)\frac{\mathrm d}{\mathrm
  dt}\left(t^4v\right)+\frac{\mathrm d^2}{\mathrm dt^2}\left(t^2v\right)+\mathrm{Ri}_cv=0\hbox{,}
\end{equation}
and
\begin{equation}
\label{Eq:Conditionab}
\left[-\left(1+\frac{1}{P_{\mathrm r}}\right)t^4ve^{\eta t}-\frac{\mathrm d}{\mathrm dt}\left(t^2v\right)e^{\eta t}+zt^2ve^{\eta t}\right]_a^b=0\hbox{.}
\end{equation}
The new variable $u$ defined by $v=t^{-2}e^{-\frac{1}{3}t^3}u$ transforms the original equation into
\begin{eqnarray}
\frac{\mathrm d^2u}{\mathrm dt^2}-\left(1-\frac{1}{P_{\mathrm r}}\right)t^2\frac{\mathrm du}{\mathrm dt}+\frac{l(l+1)}{m^2}\mathrm{Ri}_c\frac{1}{t^{2}}u=0 \hbox{.}
\end{eqnarray}

Then, we introduce $s=Dt^3$ where $D\in\mathbb{C}^{*}$: 
\begin{eqnarray}
9s^2\frac{\mathrm d^2u}{\mathrm ds^2}+\left[6s-\frac{3s^2}{D}\left(1-\frac{1}{P_{\mathrm r}}\right)\right]\frac{\mathrm du}{\mathrm ds}+\frac{l(l+1)}{m^2}\mathrm{Ri}_cu=0\hbox{.}
\end{eqnarray}

Finally, $u=s^{-1/3}Ve^{\frac{1}{6D}\left(1-\frac{1}{P_{\mathrm r}}\right)s}$ leads to: 
\begin{eqnarray}
9s^2\frac{\mathrm d^2V}{\mathrm ds^2} +&&\left[2+\frac{l(l+1)}{m^2}\mathrm{Ri}_c+
\frac{s}{D}\left(1-\frac{1}{P_{\mathrm r}}\right)\right. \\
&&-\left.\frac{s^2}{4D^2}\left(1-\frac{1}{P_{\mathrm r}}\right)^2\right]V=0
\end{eqnarray}

To obtain the final equation, we define: 
\begin{eqnarray}
&&D=-\frac{1}{3}(1-\frac{1}{P_{\mathrm r}}) \hbox{,}\\
&&M_{l,m}^2=\frac{1}{4}-\frac{2+\frac{l(l+1)}{m^2}\mathrm{Ri}_c}{9} \hbox{,}\\
&&\Lambda=-\frac{1}{3} \hbox{.}\\
\end{eqnarray}
and we get the following Whittaker equation: 
\begin{equation}
\label{Eq:Whittaker}
\frac{\mathrm d^2V}{\mathrm ds^2}
+\left(\frac{\frac{1}{4}-M_{l,m}^2}{s^2}+\frac{\Lambda}{s}-\frac{1}{4}\right)V=0 \hbox{.}
\end{equation}

\bibliographystyle{aa}  
\bibliography{AMDbiblio} 
 
\end{document}